\documentclass[twocolumn]{autart}    

\usepackage{graphicx}          
\usepackage{amssymb, amsmath,amsfonts,verbatim, mathtools,enumerate,algorithm,dsfont}
\usepackage{comment}
\usepackage{diagbox}
\usepackage{times,algpseudocode}
\usepackage[style=base]{subcaption}

\allowdisplaybreaks[2]
\newcommand{\ones}{\mathbf 1}
\newcommand{\reals}{{\mathbb{R}}}
\newcommand{\naturals}{{\mathbb{N}}}

\newcommand{\intervalSet}[2][]{[\underline{#2}_{#1}, \overline{#2}_{#1}]}

\newcommand{\argmin}{\mathop{\rm argmin}}

\newcommand{\norm}[1]{\left\lVert#1\right\rVert}
\newcommand{\mnorm}[1]{{\left\vert\kern-0.25ex\left\vert\kern-0.25ex\left\vert #1 
    \right\vert\kern-0.25ex\right\vert\kern-0.25ex\right\vert}}

\newcommand{\mc}{\mathcal}

\algnewcommand\algorithmicforeach{\textbf{for each}}
\algdef{S}[FOR]{ForEach}[1]{\algorithmicforeach\ #1\ \algorithmicdo}

\newcommand{\cl}{\operatorname{cl}}
\newcommand{\diam}[1]{\operatorname{diam}\left(#1\right)}
\newcommand{\qedwhite}{\hfill \ensuremath{\Box}}
\usepackage[dvipsnames]{xcolor}

 \newcommand\ForAuthorsAssa[1]
 {\par\smallskip                     
  \begin{center}
  \fbox
  {\parbox{0.9\linewidth}
    {\raggedright\sc--- {\color{blue}#1}}
  }
  \end{center}
  \par\smallskip                     %
 }
\begin{document}

\begin{frontmatter}
\title{Bounding Fixed Points of Set-based Bellman Operator and Nash Equilibria of Stochastic Games} 


\author[SL]{Sarah H.Q. Li},  
\author[AA]{Assal\'{e} Adj\'{e}},  
\author[Third]{Pierre-Lo\"{i}c Garoche}, 
\author[SL]{Beh\c cet A\c c\i kme\c se}

\address[SL]{William E.Boeing Department of Aeronautics and Astronautics, University of Washington, Seattle, USA. (e-mail:\{sarahli, behcet\}@uw.edu).}
\address[AA]{LAMPS, Université de Perpignan Via Domitia, Perpignan, France
 (e-mail: assale.adje@univ-perp.fr)}
\address[Third]{ENAC, Université de Toulouse, Toulouse, France  (e-mail: Pierre-Loic.Garoche@enac.fr)}

\begin{abstract}                
Motivated by uncertain parameters encountered in Markov decision processes (MDPs) and stochastic games, we study the effect of parameter uncertainty on Bellman operator-based algorithms under a set-based framework.
Specifically, we first consider a family of MDPs where the cost parameters are in a given compact set; we then define a Bellman operator acting on a set of value functions to produce a new set of value functions as the output under all possible variations in the cost parameter. We prove the existence of a {\em fixed point} of this set-based Bellman operator by showing that it is contractive on a complete metric space, and explore its relationship with the corresponding family of MDPs and stochastic games. Additionally, we show that given interval set-bounded cost parameters, we can form exact bounds on the set of optimal value functions. Finally, we utilize our results to bound the value function trajectory of a player in a stochastic game.
\end{abstract}


\end{frontmatter}

\section{Introduction}
Markov decision process (MDP) is a fundamental framework for control design in stochastic environments, reinforcement learning, and stochastic games~\cite{accikmecse2015markov,demir2015decentralized,filar2012competitive,li2019tolling}. Given cost and transition probabilities, solving an MDP is equivalent to minimizing an objective in expectation, and requires determining the optimal value function as well as deriving the corresponding optimal policy for each state. Relying on the fact that the optimal value function is the \emph{fixed point} of the Bellman operator, dynamic programming methods iteratively apply variants of the Bellman operator to converge to the optimal value function and the optimal policy~\cite{puterman2014markov}.

We are motivated to study MDPs where the parameters that define the environment are \emph{sets} rather than single-valued. Such a set-based perspective arises naturally in the analysis of parameter uncertain MDPs and stochastic games. In this paper, we develop a framework for evaluating MDPs on compact sets of costs and value functions. Specifically, we show that when the cost parameter of the MDP is in a compact set rather than single-valued, we can define a Bellman operator on the space of compact sets, such that it is contractive with respect to the Hausdorff distance. 
We prove the existence of a unique and compact \emph{fixed point set} that the operator must converge to, and give interpretations of the fixed point set in the context of parameter uncertain MDPs and stochastic games.  

When modeling a system as a stochastic process, sampling techniques are often used to determine cost and transition probability parameters. In such scenarios, the MDP can be either interpreted as a standard MDP with error bounds on its parameters, or as a \emph{set-based MDP} in which its parameters are sets rather than single-valued. In the former approach, an MDP can be solved with standard dynamic programming methods, and the stability of its solution with respect to parameter perturbation can be analyzed locally~\cite{abbad1992perturbation,altman1993stability,bielecki1991singularly}. However, these sensitivity results are only local approximations in the context of compact parameter sets. The latter approach is not well explored --- some research exists on bounded interval set MDPs~\cite{givan2000bounded}, in which dynamic programming techniques such as value  and policy iteration have been shown to converge. However, while it is known that parameter uncertain MDPs may result value function sets such as polytopes~\cite{DBLP:conf/icml/DadashiBTRS19}, there is no convergence guarantees for dynamic programming with polytopic sets of value functions. In this paper, we show that for a set-based MDP with a compact set of cost parameters, and any regular MDP whose cost is an element of the said compact set of cost parameters, the associated set-based Bellman operator has a unique and compact fixed point set that must contain the optimal value function of the regular MDP.

As opposed to parameter uncertain MDPs where the underlying cost and probability parameters are constant albeit uncertain, stochastic games result in MDPs where the cost and probability parameters \emph{vary} with opponents' changing policies. An individual player can interpret a stochastic game as an MDP with a  parameter-varying environment,  
where holding all opponents' policies fixed, the stochastic game played by player $i$ is equivalent to a regular MDP. At a fixed joint policy, we say that a player's policy is \emph{optimal} if it is optimal with respect to the corresponding MDP. If every player's policy is optimal with respect to their opponents' fixed policies within a joint policy, then we say that the game has reached a Nash equilibrium --- i.e., every player's policy is \emph{optimal} for the current joint policy. A Nash equilibrium defines a joint policy at which no player has any incentive to deviate from.  
In learning theory for stochastic games, it is often each player's goal to achieve the Nash equilibrium through an iterative process. Therefore, many learning algorithms are based on variants of dynamic programming, where each player solves an MDP with costs and transition probabilities changing at each iteration~\cite{bu2008comprehensive,littman2001value}. 
In this paper, we apply a set-based dynamic programming technique to single controller stochastic games. However, rather than demonstrating convergence toward a Nash equilibrium, we show that the \emph{set} of Nash equilibria must be contained in the fixed point set of a set-based Bellman operator.

In~\cite{li2020setBellman}, we began our analysis of set-based MDPs by proving the existence of a unique fixed point set associated to the set-based Bellman operator. In this paper, we demonstrate the significance of this fixed point set by relating it to the fixed points of parameter uncertain MDPs and the Nash equilibria set of stochastic games. We further explore the fixed point set in the context of iterative solutions to stochastic games, and show that the fixed point set of the set-based Bellman operator bounds the asymptotic behaviour of dynamic programming-based learning algorithms. 

The paper is structured as follows: we provide references to existing research in Section~\ref{sec:relatedWorks}; we recall definitions of an MDP and the Bellman operator in Section~\ref{sec:setup}; Section~\ref{sec:set-VI} extends these definitions to set-based MDPs, providing theoretical results for the existence of a fixed point set of a set-based Bellman operator. Section~\ref{sec:connection2StochasticGames} relates properties of the fixed point set to stochastic games. An interval set-based MDP is presented in Section~\ref{sec:ex} with a computation of exact bounds, while the application to stochastic games is illustrated in Section~\ref{sec:stochasticGames}, where we model unknown policies of the opponent as cost intervals.

\section{Related Research}
\label{sec:relatedWorks}

Bounding the fixed point of the Bellman operator with uncertain parameters is well studied under robust MDPs such as in~\cite{delage2010percentile,wiesemann2013robust}, where the MDP parameters are either assumed or estimated as random variables from a known Gaussian distribution~\cite{delage2010percentile,wiesemann2013robust}. In contrast, we model our MDP cost parameter uncertainty as a compact set without any probabilistic prior assumptions. Therefore, our results are absolute as opposed to chanced constrained or stochastic.

A closely related work from robust MDP is~\cite{iyengar2005robust}, where the author analyzes what we consider as the lower bound on the value function of parameter uncertain MDPs
and connects parameter uncertain MDPs to perfect information stochastic games. 
We generalize these results for cost uncertainty only, and show that there exists an invariant set corresponding to parameter uncertain MDPs, and that dynamic programming-based algorithms can converge to the invariant set itself instead of just obtaining a lower bound. 

Our parameter uncertain MDP model also generalizes the bounded parameter model presented in~\cite{givan2000bounded}, which considers interval sets instead of general compact sets.

Other approaches to bound the value functions of cost uncertain MDPs include~\cite{dick2014online,haddad2018interval}. MDPs with reachability objectives is studied in~\cite{haddad2018interval} under a graph theoretical MDP uncertainty model. However, the techniques utilized in~\cite{haddad2018interval} require abstraction of the MDP state space and therefore do not directly extend to value functions which are defined per state. A learning approach to solve cost-uncertain MDPs is taken in~\cite{dick2014online}, and convergence in terms of regret is shown using gradient-based algorithms instead of dynamic programming approaches.

Introduced in~\cite{shapley1953stochastic}, stochastic games generalizes MDPs to the multi-agent setting, where the goal of each player is to achieve a Nash equilibrium.  
In general, it is difficult to find a Nash equilibrium of a general sum stochastic game; the computation complexity has been shown to be NP hard in~\cite{chatterjee2004nash}, while value iteration for such games is shown to diverge in~\cite{kearns2000fast}. Convergence guarantees for Bellman operator-based algorithms exist under limited settings such as two player stochastic games or zero sum stochastic games~\cite{shapley1953stochastic,eisentraut2019stopping,prasad2015two,wei2017online}. However, the same algorithms have been shown to converge empirically in a wide range of applications including poker and cyber-security~\cite{ganzfried2009computing,shiva2010game}.
In this paper, we consider single controller stochastic games with \emph{imperfect information}~\cite[Def. 6.3.6]{filar2012competitive}, and show that our set-based value iteration algorithm converges to an invariant set that over-approximates the Nash equilibrium set. 

The topology of value function sets has also garnered interest in the reinforcement learning community~\cite{bellemare2019geometric,DBLP:conf/icml/DadashiBTRS19}. In~\cite{DBLP:conf/icml/DadashiBTRS19}, the set of value functions generated by policy uncertainty is shown to be a polytope, and Bellman operator-based methods such as value iteration and policy iteration are shown to converge to the value function polytope.

\section{MDP and Bellman Operator}
\label{sec:setup}
We introduce our notation for existing results in MDP literature, which is used throughout the paper. Contents from this section are discussed in further detail in~\cite{puterman2014markov}. 

\textbf{Notation}: Sets of $N$ elements are given by $[N] = \{0, \ldots, N-1\}$. We denote the set of matrices with $i$ rows and $j$ columns of real or non-negative valued entries as $\reals^{i \times j}$  or $\reals_+^{i\times j}$, respectively. Matrices and some integers are denoted by capital letters, $X$, while sets are denoted by cursive letters, $\mc{X}$. The set of all \emph{non-empty compact subsets} of $\mc{X}$ is denoted by $H(\mc{X})$. The column vector of ones is denoted by $\ones_N = [1, \ldots, 1]^\top \in \reals^{N\times 1}$. The identity matrix of size $S\times S$ is denoted by $I_S$.

We consider a \emph{discounted infinite-horizon MDP} defined by $([S], [A], P, C, \gamma)$, where
\begin{enumerate}
	\item $[S]$ denotes the finite set of states.
	\item $[A]$ denotes the finite set of actions. Without loss of generality, assume that every action is admissible from each state $s \in [S]$.

	\item $P \in \reals^{S \times SA}$ defines the transition kernel. Each component $P_{s',sa}$ is the probability of arriving in state $s'$ by taking state-action $(s,a)$. Matrix $P$ is column stochastic and element-wise non-negative --- i.e.,
	\begin{equation}
	    \begin{aligned}
	        \underset{s' \in [S]}{\sum} P_{s',sa} = 1, \ \forall \ (s,a) \in [S]\times [A], \\
	        P_{s',sa} \geq 0 , \ \forall \ (s',s, a) \in [S]\times[S]\times[A].
	    \end{aligned}
	\end{equation}
	\item $C \in \reals^{S \times A}$ defines the cost matrix. Each component $C_{sa}$ is the cost of state-action pair $(s,a) \in [S] \times [A]$.
	\item $\gamma \in (0,1)$ denotes the discount factor.  
\end{enumerate}

At each time step $t$, the decision maker chooses an action $a$ at its current state $s$. The state-action pair $(s, a)$ induces a probability distribution vector over states $[S]$ as $[P_{1,sa}, P_{2,sa}, \ldots, P_{S, sa}]$. The state-action $(s,a)$ also induces a cost $C_{sa}$ for the decision maker. 

The decision maker chooses actions via a \emph{policy}. We denote a policy as a function $\pi:\reals^{S}\times\reals^{A} \rightarrow [0,1]$, where $\pi(s,a)$ denotes the probability that action $a$ is chosen at state $s$. 
The set of all policies of an MDP is denoted by ${\Pi}$. Within $\Pi$, a policy $\pi$ is deterministic if at each state $s$, $\pi(s,a)$ returns $1$ for exactly one action, and $0$ for all other possible actions. A policy $\pi \in \Pi$ that is not deterministic is a \emph{mixed policy}. 

We denote the policy matrix induced by the policy $\pi$ as $M_\pi \in \reals^{S\times SA}$, where 
\begin{equation}
(M_\pi)_{s', sa} = \begin{cases}
\pi(s,a) & s' = s \\
0 & s' \neq s
\end{cases}.
\end{equation}
Every policy induces a \emph{Markov chain}~\cite{el2018controlled}, given by $M_{\pi}P^\top \in \reals^{S\times S}$. 
Each stationary policy also induces a stationary cost given by
\begin{equation}\label{eqn:policyCost}
  \nu(\pi)= \sum_{i \in [S]} e_i e_i^\top M_{\pi}(\ones_S \otimes I_A) C^\top e_i, \ \nu(\pi) \in \reals^{S}, 
\end{equation}
where $e_i \in \reals^{S}$ is the unit vector pointing in the $i^{th}$ coordinate, $\otimes$ is the Kronecker product, and $I_A$ is the identity matrix of size $A$.


For an MDP $([S],[A],P,C,\gamma)$, we are interested in minimizing the \emph{discounted infinite horizon expected cost}, defined with respect to a policy $\pi$ as
\begin{equation}\label{eqn:expectationOptimalV}
 V^\star_{s} =\ \min_{\pi \in \Pi} \ \mathbb{E}^\pi_{s} \Big\{ \sum_{t = 0}^\infty \gamma^t C_{s^t a^t}\Big\}, \quad \forall \ s \in [S],
\end{equation}
where $\mathbb{E}_{s}^\pi(f)$ is the discounted infinite horizon expected value of objective $f$, $s^t$ and $a^t$ are the state and action taken at time step $t$, and $s$ is the initial state of the decision maker at $t = 0$.  

$V_s^\star$ is the \emph{optimal value function} for the initial state $s$. The policy $\pi^\star$ that achieves this optimal value is called an \emph{optimal policy}. In general, the optimal value function $V^\star_s$ is unique while the optimal policy $\pi^\star$ is not. The set of optimal policies always includes at least one deterministic stationary policy in the unconstrained setting~\cite[Thm 6.2.11]{puterman2014markov}. If there are constraints on the policy and state space, deterministic policies may become infeasible~\cite{el2018controlled}.  

\subsection{Bellman Operator}
Determining the optimal value function of a given MDP is equivalent to solving for the fixed point of the associated Bellman operator, for which a myriad of techniques exists~\cite{puterman2014markov}. We introduce the Bellman operator and its fixed point here for the corresponding MDP problem.

\begin{defn}[Bellman Operator]\label{def:bellmanOp} 
For a discounted infinite horizon MDP $([S],[A],P,C,\gamma)$, its Bellman operator $f_C: \reals^{S}\rightarrow \reals^{S}$ is given component-wise as 
\begin{equation}
\Big(f_C(V)\Big)_s :=  \min_{a}\ C_{sa} +\gamma\sum_{s'\in[S]} P_{s',sa}V_{s'},\ \forall\, s\in [S].\label{eqn:bellmanOp}    
\end{equation}
\end{defn}
The fixed point of the Bellman operator is a value function $V \in \reals^S$ that is invariant with respect to the operator. 
\begin{defn}[Fixed Point]\label{def:fixedPoint}
$V^\star$ is a fixed point of an operator $F: \mc{X} \mapsto \mc{X}$ iff
\begin{equation}
V^\star = F(V^\star).
\end{equation}
\end{defn}
In our discussion of the fixed point of the Bellman operator, we consider the following operator properties.


\begin{defn}[Order Preservation]\label{def:orderPreserve}
Let $\mc{X}$ be a partially ordered space with partial order $\preceq$. An operator $F:\mc{X}\rightarrow \mc{X}$ is an order preserving operator iff  
\[x \preceq x' \rightarrow\ F(x) \preceq F(x'), \quad \forall \ x, x' \in \mc{X}. \]
\end{defn}
\begin{defn}[Contraction]\label{def:contractOp}
Let $(\mc{X}, d)$ be a complete metric space with metric $d$. An operator $F:\mc{X}\mapsto \mc{X}$ is a contracting operator iff
\[d(F(x), F(x')) < d(x, x'), \quad \forall \ x, \ x' \ \in \mc{X}.\]
\end{defn}
The Bellman operator $f_C$ is known to have both properties on the complete metric space $(\reals^{S}, \norm{\cdot}_\infty)$. Therefore, the Banach fixed point theorem can be used to show that $f_C$ has a unique fixed point~\cite{puterman2014markov}. 
Because the optimal value function $V^\star$ is given by the unique fixed point of the associated Bellman operator $f_C$, we use the terms optimal value function and fixed point of $f_C$ interchangeably. 

In addition to obtaining $V^\star$, MDPs are also solved to determine the \emph{optimal policy}, $\pi^\star$. Every policy $\pi$ induces a unique \emph{stationary} value function $V$ given by
\begin{equation}\label{eqn:vecForm_of_VI}
    V = \nu(\pi)+ \gamma M_{\pi}P^\top V,
\end{equation}
where $\gamma \in (0,1)$. We note that $V$ is a linear function of $C$ through $\nu(\pi)$ in~\eqref{eqn:policyCost}, where the dependency is made implicit to simplify notation.

Given a policy $\pi$, we can equivalently solve for the stationary value function $V$ as $V = (I - \gamma M_{\pi}P^\top )^{-1}\nu(\pi)$. From this perspective, the optimal value function $V^\star$ is the minimum vector among the set of stationary value functions corresponding to the set of policies $\Pi$. 
Policy iteration algorithms utilize this fact to obtain the optimal value function $V^\star$ by iterating over the feasible policy space~\cite{puterman2014markov}.  

Given an input value function $V$, we can also derive a deterministic optimal policy $\pi$ associated with $f_C(V)$ as
\begin{equation}\label{eqn:optimalPol}
    \pi(s, a) := \begin{cases} 1 & a = \underset{a' \in [A]}{\argmin} \ C_{sa'} + \gamma \underset{s'\in[S]}{\sum}P_{s',sa'}V_{s'} \\
    0 & \text{otherwise} 
    \end{cases},
\end{equation}
where $\argmin_{a'\in[A]}$ returns the first optimal action $a'$ if multiple actions minimize the expression $C_{sa'} + \gamma{\sum}_{s'\in[S]}P_{s',sa'}V_{s'}$ at state $s$. 



While policies that solve $f_C(V)$ do not need to be unique, deterministic or stationary, the policy $\pi$ derived from~\eqref{eqn:optimalPol} will always be unique, deterministic and stationary for a given ordering of actions within the action set. For the remaining sections, we assume that the action set $[A]$ have a fixed ordering of actions.

\subsection{Termination Criteria for Value Iteration}
Among different algorithms that solve for the fixed point of the Bellman operator, \emph{value iteration} (VI) is a commonly used and simple technique in which the Bellman operator is iteratively applied until the optimal value function is reached --- i.e. starting from any value function $V^0 \in \reals^S$, we apply
\begin{equation} \label{eqn:VI}
    \begin{array}{ll}
    V^{k+1}_s &=f_C(V^k)\\
    &= \underset{a\in[A]}{\min} \ C_{sa} + \gamma \underset{s'\in [S]}{\sum} P_{s',sa}V^k_{s'}, \quad k = 1, 2, \ldots.
    \end{array}
    \end{equation}


The iteration scheme given by~\eqref{eqn:VI} converges to the optimal value function of the corresponding discounted infinite horizon MDP. 
The following result presents a \emph{stopping criteria} for~\eqref{eqn:VI}. 
\begin{lem}\label{thm:stoppingCriteria}\cite[Thm. 6.3.1]{puterman2014markov}
For any initial value function $V^0 \in \reals^S$, let $\{V^k\}_{k\in \naturals}$ satisfy the value iteration given by~\eqref{eqn:VI}. For $\epsilon > 0$, if
\[\norm{V^{k+1} - V^k}_{\infty} < \epsilon \frac{(1 - \gamma)}{2\gamma},\]
then $V^{k+1}$ is within $\epsilon /2$ of the fixed point $V^\star$, i.e. 
\[\norm{V^{k+1} - V^\star}_{\infty} < \frac{\epsilon}{2}.\]
\end{lem}
Lemma~\ref{thm:stoppingCriteria} connects relative convergence of the sequence $\{V^k\}_{k \in \naturals}$ to absolute convergence towards $V^\star$ by showing that the former implies the latter. In general, the stopping criteria differ for different MDP objectives (see~\cite{haddad2018interval} for recent results on stopping criteria for MDPs with a reachability objective).

\section{Set-based Bellman Operator}
\label{sec:set-VI}
The classic Bellman operator with respect to a cost $C$ is well studied. Motivated by parameter uncertain MDPs and stochastic games, we extend the classic Bellman operator by \emph{lifting} it to operate on sets rather than individual value functions in $\reals^S$. For the set-based operator, we analyze its set-based domain and prove relevant operator properties such as order preservation and contraction. Finally, we show the existence of a unique fixed point \emph{set} $\mc{V}^\star$ and relate its properties to the fixed point of the classic Bellman operator.  
\subsection{Set-based operator properties}
For the domain of our set-based operator, we define a new metric space $(H(\reals^{S}), d_H)$ based on the Banach space $(\reals^S, \norm{\cdot}_\infty)$ \cite{rudin1964principles}, where $H(\reals^{S})$ denotes the collection of non-empty compact subsets of $\reals^{S}$. We equip $H(\reals^{S})$ with \emph{partial order}, $\preceq$, where for $\mc{V}, \mc{V}' \in H(\reals^S)$, $\mc{V} \preceq \mc{V}'$ iff $\mc{V} \subseteq \mc{V}'$. The metric $d_H$ is the  following \emph{Haussdorf distance}~\cite{henrikson1999completeness} defined as 
\begin{align}
d_H(\mc{V}, \mc{V}') := \max\{ &\sup_{V \in \mc{V}}\inf_{V' \in \mc{V}'} \norm{V - V'}_\infty,\\
&\sup_{V' \in \mc{V}'}\inf_{V \in \mc{V}} \norm{V - V'}_\infty \}.
\end{align}

\begin{lem}\cite[Thm 3.3]{henrikson1999completeness}\label{lem:hausdorffComplete}
If $\mc{X}$ is a complete metric space, then its induced Hausdorff metric space $(H(\mc{X}), d_H)$ is a complete metric space.
\end{lem}
From Lemma~\ref{lem:hausdorffComplete}, since $(\reals^S, \norm{\cdot}_\infty)$ is a complete metric space, $H(\reals^{S})$ is a complete metric space with respect to $d_H$. 
On the complete metric space $H(\reals^{S})$, we define a \emph{set-based Bellman operator} which acts on compact sets. 
\begin{defn}[Set-based Bellman Operator]\label{def:setBasedBellman}
For a family of MDP problems, $([S], [A], P, \mc{C}, \gamma)$, where $\mc{C} \subset \reals^{S\times A}$ is a non-empty compact set, its associated set-based Bellman operator is given by
\[F_{\mc{C}}(\mc{V}) := \cl\bigcup_{(C, V) \in\mc{C}\times \mc{V}} f_C(V), \quad \forall \ \mc{V} \in H(\reals^S),\]
where $\cl$ is the closure operator.
\end{defn}
Since $F_{\mc{C}}$ is the union of uncountably many bounded sets, the resulting set may not be bounded, and therefore it is not immediately obvious that $F_{\mc{C}}(\mc{V})$ maps into the metric space $H(\reals^S)$. 
\begin{prop}\label{prop:BellmanCompact}
If $\mc{C}$ is non-empty and compact, then $F_{\mc{C}}(\mc{V})\in H(\reals^{S})$, $\forall \ \mc{V}\in H(\reals^{S})$.
\end{prop}
\begin{pf}
For a non-empty and bounded subset $\mc{A}$ of a finite dimensional real vector space, we define its diameter as $\diam{\mc{A}}=\sup_{x,y\in \mc{A}} \norm{x-y}_\infty$. The diameter of a set in a metric space is finite if and only if it is bounded~\cite{rudin1964principles}.

Take any non-empty compact set $\mc{V}\in H(\reals^{S})$. As $F_{\mc{C}}(\mc{V})\subseteq \reals^S$, it suffices to prove that $F_{\mc{C}}(\mc{V})$ is closed and bounded. The closedness is guaranteed by the closure operator. A subset of a metric space is bounded iff its closure is bounded. Hence, to prove the boundedness, it suffices to prove that $\diam{\cup_{(C,V) \in\mc{C}\times\mc{V}} f_C(V)}<+\infty$. 
For any two cost-value function pairs $(C,V), (C',V') \in \mc{C} \times \mc{V}$,
\begin{equation}\label{eqn:pf_compactOperator_1}
   f_{C}(V) - f_{C'}(V') = \Big(f_{C}(V) - f_{C'}(V)\Big) + \Big(f_{C'}(V) - f_{C'}(V')\Big). 
\end{equation}
We bound~\eqref{eqn:pf_compactOperator_1} by bounding the two terms on the right hand side separately. The second term satisfies \(\norm{f_{C'}(V) - f_{C'}(V')}_\infty \leq \gamma\norm{V - V'}_\infty,\) due to contraction properties of $f_{C'}$. To bound the first term, we note that for any two vectors $a, b \in \reals^{S}$, 
\begin{equation}\label{eqn:infNorm2Max}
  \norm{a- b}_\infty = \max\Big\{\max(a - b), \max(b - a)\Big\},  
\end{equation}
where the operator $\max\{\ldots\}$ returns the maximum element, and $\max(a)$ returns maximum component of vector $a$. Evaluating $f_{C'}(V) - f_{C}(V)$ with~\eqref{eqn:infNorm2Max}, 
\begin{align*}
 & \max(f_{C'}(V) - f_C(V)) \\
 \leq & \max(\nu'(\pi) + \gamma M_{\pi}P^\top V - \nu(\pi)-\gamma M_{\pi}P^\top V) \\ 
 \leq & \max\Big(\nu'(\pi) - \nu(\pi)\Big) \\
 \leq& \sum_{i \in [S]}\norm{e^\top_i}_{\infty} \norm{M_{\pi}}_{\infty}\norm{ \ones_S\otimes I_A}_{\infty} \norm{(C' - C)^\top}_\infty\norm{e_i}^2_{\infty},
\end{align*}
where $\pi$ is an optimal policy corresponding to $f_{C}$.
Since $\norm{\ones_S\otimes I_A}_\infty = \norm{e_i}_\infty = \norm{e^\top_i}_\infty = \norm{M_{\pi}}_\infty = 1$ for any $\pi \in \Pi$, $ \max(f_{C'}(V) - f_C(V)) \leq  S \norm{(C' - C)^\top}$.  
Similarly, we can show $\max(f_{C}(V) - f_{C'}(V)) \leq S \norm{(C' - C)^\top}_{\infty}$.
Finally it follows from~\eqref{eqn:pf_compactOperator_1} that
\begin{equation}\label{eqn:fCisBounded}
  \norm{f_C(V) - f_{C'}(V')}_\infty \leq S \norm{(C' - C)^\top}_\infty  + \gamma \norm{V - V'}_\infty.  
\end{equation}
Since~\eqref{eqn:fCisBounded} holds for all $(C,V), (C', V') \in \mc{C} \times \mc{V}$, and furthermore, for all $C, C' \in \mc{C}$ and $V, V' \in \mc{V}$, 
\[\norm{(C' - C)^\top}_\infty \leq \diam{\mc{C}^\top}, \ \norm{V - V'}_\infty \leq \diam{\mc{V}}, \]
the inequality $\diam{\cup_{(C,V) \in\mc{C}\times\mc{V}} f_C(V)}\leq S\diam{\mc{C}^\top}+\gamma\diam{\mc{V}}<+\infty$ holds as both $\mc{C}^\top$ and $\mc{V}$ are bounded. \qedwhite
\end{pf}
Proposition~\ref{prop:BellmanCompact} shows that $F_{\mc{C}}$ is an operator from $H(\reals^S)$ to $H(\reals^S)$. Having established the space which $F_{\mc{C}}$ operates on, we can draw many parallels between $F_{\mc{C}}$ and $f_C$. Similar to $f_C$ having a unique fixed point $V^\star$ in the real vector space, we consider whether a unique \emph{fixed point set} $\mc{V}^\star$ which satisfies $F_{\mc{C}}(\mc{V}^\star) = \mc{V}^\star$ exists. To take the comparison further, since $V^\star$ is optimal for an MDP problem defined by $([S],[A],P,C,\gamma)$, we consider if $\mc{V}^\star$ correlates to the \emph{family} of optimal value functions that correspond to the MDP family $([S],[A],P,\mc{C}, \gamma)$. We explore these parallels in this paper and derive sufficient conditions for the existence and uniqueness of the fixed point of the set-based Bellman operator $F_{\mc{C}}$. 

We demonstrate the existence and uniqueness of $\mc{V}^\star$ by utilizing the Banach fixed point theorem~\cite{puterman2014markov}, which states that a unique fixed point must exist for all contraction operators on complete metric spaces. First, we show that $F_{\mc{C}}$ has properties given in Definitions~\ref{def:orderPreserve} and~\ref{def:contractOp} on the complete metric space $(H(\reals^S), d_H)$.

\begin{prop}\label{prop:setBellman}
For any $\mc{V} \in H(\reals^S)$ and $\mc{C} \subset \reals^{S\times A}$ closed and bounded, $F_{\mc{C}}$ is an order preserving and a contracting operator in the Hausdorff distance. 
\end{prop}

\begin{pf}
Consider $\mc{V}$, ${\mc{V}'} \in H(\reals^S)$ which satisfy $\mc{V} \subseteq \mc{V}'$, then 
\[F_{\mc{C}}(\mc{V}) = \cl{ \underset{\substack{(C, V) \\\in  \mc{C} \times\mc{V}}}{\bigcup} f_C(V)} \subseteq \cl{ \underset{\substack{(C, V') \\\in  \mc{C} \times\mc{V}'}}{\bigcup}f_C(V')} = F_{\mc{C}}(\mc{V}').\]
We conclude that $F_{\mc{C}}$ is order-preserving.
To see that $F_{\mc{C}}$ is contracting, we need to show 
\begin{align}\label{eqn:pf_contractingBO0}
   \sup_{V \in F_{\mc{C}}(\mc{V})}\inf_{V' \in F_{\mc{C}}(\mc{V'})} \norm{V - V'}_\infty & < d_H(\mc{V}, \mc{V}') \\
   \label{eqn:pf_contractingBO1}
   \sup_{V' \in F_{\mc{C}}(\mc{V'})} \inf_{V \in F_{\mc{C}}(\mc{V})} \norm{V - V'}_\infty & < d_H(\mc{V}, \mc{V}'), 
\end{align}
First we note that taking $\sup$ ($\inf$) of a continuous function over the closure of a set $\mc{A}$ is equivalent to taking the $\sup$ ($\inf$) over $\mc{A}$ itself. Furthermore, the single-cost Bellman operator $f_C(V)$ is an element of the set-based Bellman operator $\cup_{(C,V) \in \mc{C} \times \mc{V}} f_C(V)$ iff $(C,V) \in \mc{C} \times \mc{V}$. 
Therefore taking the $\sup(\inf)$ of $\norm{V -V'}_\infty$ over $V\in F_{\mc{C}}(\mc{V})$ is equivalent to taking the $\sup(\inf)$ of $\norm{f_C(V) - f_{C'}(V')}_\infty$ over $(C,V) \in \mc{C}\times \mc{V}$. 

Given $\mc{V}, \mc{V}' \in H(\reals^S)$ and for arbitrary $V \in \mc{V}$,  $C\in\mc{C}$,
\begin{subequations}
\begin{align}
\label{prf:contraction0}
& \underset{\substack{(C', V') \\\in  \mc{C} \times\mc{V}'}}{\inf} \norm{f_C(V) - f_{C'}(V')}_\infty\\
\label{prf:contraction1}
\leq & \underset{\substack{(C', V') \\\in  \mc{C} \times\mc{V}'}}{\inf}  S\norm{(C - C')^\top}_\infty + \gamma \norm{V - V'}_\infty\\
\label{prf:contraction3}
\leq & S\norm{(C' - C')^\top}_\infty +  \underset{\substack{V' \in  \mc{V}'}}{\inf} \gamma\norm{V - V'}_\infty \\
\label{prf:contraction4}
\leq & \gamma \underset{\substack{V' \in  \mc{V}'}}{\inf} \norm{V - V'}_\infty,
\end{align}
\end{subequations}
where in~\eqref{prf:contraction1} we take the upper bound derived in~\eqref{eqn:fCisBounded}. In~\eqref{prf:contraction3} we haven chosen the matrix $C= C'$ to minimize $\norm{(C - C')^\top}_{\infty}$. This eliminates the cost term and we arrive at~\eqref{prf:contraction4}. 

Then~\eqref{eqn:pf_contractingBO0} and~\eqref{eqn:pf_contractingBO1} simplifies to 
\[\sup_{V \in F_{\mc{C}}(\mc{V})}\inf_{V' \in F_{\mc{C}}(\mc{V}')} \norm{V - V'}_\infty \leq \gamma \sup_{V \in \mc{V}}\inf_{V' \in \mc{V}'} \norm{V - V'}_\infty,\] and \[\sup_{V' \in F_{\mc{C}}(\mc{V}')}\inf_{V \in F_{\mc{C}}(\mc{V})} \norm{V - V'}_\infty \leq \gamma \sup_{V' \in \mc{V}'}\inf_{V \in \mc{V}} \norm{V - V'}_\infty.\] 
Therefore
$d_H(F_{\mc{C}}(\mc{V}), F_{\mc{C}}(\mc{V}')) \leq \gamma d_H(\mc{V}, \mc{V}')$.
Since $\gamma \in (0,1)$, $F_{\mc{C}}$ is a contracting operator on $H(\reals^S)$. \qedwhite
\end{pf}

The contraction property of $F_{\mc{C}}$ implies that any repeated application of the operator to a set $\mc{V}^0 \in H(\reals^S)$ results in a sequence of sets where consecutive sets become increasingly closer in the Hausdorff distance. It is then natural to consider if there is a unique set which all $F_{\mc{C}}(\mc{V}^k)$ converges to. 

\begin{thm}\label{thm:uniqueSetFixedPont}
There exists a unique fixed point $\mc{V}^\star$ of the set-based Bellman operator $F_{\mc{C}}$ as defined in Definition~\ref{def:bellmanOp}, such that $F_{\mc{C}}(\mc{V}^\star) = \mc{V}^\star$, and $\mc{V}^\star$ is a closed and bounded set in $\reals^S$.

Furthermore, for any set $\mc{V}^0 \in H(\reals^S)$, the iteration 
\begin{equation}\label{eqn:setbased_VI}
    \mc{V}^{k+1} = F_{\mc{C}}(\mc{V}^k),
\end{equation}
converges in the Haussdorf distance --- i.e.,
\[\lim_{k\to\infty} d_H(F_{\mc{C}}(\mc{V}^k),\mc{V}^\star)=0.\]
\end{thm}
\begin{pf}
As shown in Proposition~\ref{prop:setBellman}, $F_{\mc{C}}$ is a contracting operator. From the Banach fixed point theorem~\cite[Thm 6.2.3]{puterman2014markov}, there exists a unique fixed point $\mc{V}^\star$, and any arbitrary $\mc{V}^0 \in H(\reals^S)$ will generate a sequence of sets $\{F_{\mc{C}}(\mc{V}^k)\}_{k\in\naturals}$ that converges to $\mc{V}^\star$. \qedwhite
\end{pf}

\subsection{Properties of fixed point set}
In the case of the Bellman operator $f_{C}$ on metric space $\reals^S$, the fixed point $V^\star$ corresponds to the optimal value function of the MDP associated with cost $C$. Because there is no direct association of an MDP problem to the set of cost parameters $\mc{C}$, we cannot claim the same for the set-based Bellman operator and $\mc{V}^\star$. However, $\mc{V}^\star$ does have many interesting properties on $H(\reals^S)$, especially in terms of set-based value iteration~\eqref{eqn:setbased_VI}.  

We consider the following generalization of value iteration: suppose that instead of a fixed cost parameter, we have that at each iteration $k$, a $C^k$ that is randomly chosen from the compact set of cost parameters $\mc{C}$. In general, $\lim_{k\rightarrow \infty} f_{C^k}(V^k)$ may not exist. However, we can infer from Theorem~\ref{thm:uniqueSetFixedPont} that the sequence $\{V^k\}$ converges to the set $\mc{V}^\star$ in the Hausdorff distance. 

\begin{prop}\label{prop:setConverge}
Let $\{C^k\}_{k\in\naturals} \subseteq \mc{C}$ be a sequence of costs in $\mc{C}$, where $\mc{C}$ is a compact set within $\reals^{S\times A}$. Let us define the iteration
\[
V^{k+1}=f_{C^k}(V^k),
\]
for any $V^0\in\reals^S$. 
Then the sequence $\{V^k\}_{k \in \naturals}$ satisfies
\[\lim_{k \rightarrow \infty} \inf_{V\in\mc{V}^\star} \norm{f_{C^k}(V^k)-V}_\infty = 0, \] 
where $\mc{V}^\star$ is the unique fixed point set of the operator $F_{\mc{C}}$. 
\end{prop}
\begin{pf}
Define $\mc{V}^0 = \{V^0\}$, then from Definition~\ref{def:setBasedBellman} and Definition~\ref{def:bellmanOp}, $V^{k+1} = f_{C^k}(V^k) \in F_{\mc{C}}(\mc{V}^{k})$ for all $k \geq 0$. 

At each iteration $k$, we write $\mc{V}^{k+1} = F_{\mc{C}}(\mc{V}^k)$. From Theorem~\ref{thm:uniqueSetFixedPont},  $\mc{V}^k$ converges to $\mc{V}^\star$ in Hausdorff distance, $\lim_{k\rightarrow\infty}d_H(\mc{V}^k,  \mc{V}^\star) = 0$. Therefore for every $\delta > 0$, there exists $K$ such that for all $k \geq K$, $d_H(\mc{V}^k, \mc{V}^\star) < \delta$.  Since $f_{C^k}(V^k) \in \mc{V}^{k+1}$, $\inf_{V\in \mc{V}^\star} \norm{f_{C^k}(V^k)-V}_\infty \leq d_H(\mc{V}^{k+1}, \mc{V}^\star) < \delta$ must also be true for all $k \geq K$. Therefore $\lim_{k\rightarrow \infty}\inf_{V\in \mc{V}^\star} \norm{f_{C^k}(V^k)-V}_\infty = 0 $. \qedwhite
\end{pf}

Proposition~\ref{prop:setConverge} implies that regardless of whether or not the sequence $\{f_{C^k}(V^k)\}_{k\in\naturals}$ converges, the sequence $\{V^k\}$ must  become arbitrarily close in Hausdorff distance to the set $\mc{V}^\star$. This has important interpretations in the game setting that is further explored in Section~\ref{sec:connection2StochasticGames}. On the other hand, Proposition~\ref{prop:setConverge} also implies that if $\lim_{k\rightarrow \infty}V^k$ does converge, its limit point must be an element of $\mc{V}^\star$. 
\begin{cor}\label{cor:fCk_convergence}
We define the set of fixed points of $f_C$ for each $C \in\mc{C}$ as 
\[ \mc{U} = \underset{C \in \mc{C}}{\bigcup} \{ V \in \reals^{S} \ | \ f_C(V) = V \},\]
i.e., $\mc{U}$ is the set of optimal value functions for the set of MDPs $([S],[A], P, C, \gamma)$ where $C \in \mc{C}$. Furthermore, we consider all sequences $\{C^k\}_{k\in\naturals}\subseteq \mc{C}$ such that for $V^0 \in \reals^{S}$, the iteration $V^{k+1} = f_{C^k}(V^k)$ approaches a limit point  $V = \lim_{k\rightarrow\infty}V^k$, and define the set of all such limit points as 
\begin{align}
\mc{W} = & \underset{\substack{\{C^k\}_{k \in \naturals} \subseteq \mc{C} }}{\bigcup}\{ V \in \reals^{S} \ | \ \lim_{k\rightarrow \infty} f_{C^k}(V^k) = V,  \text{ where}\notag \\
& \qquad \qquad \ V^0 \in \reals^{S},  V^{k+1} =  f_{C^k}(V^k), \ k = 0, 1,\ldots\},     
\end{align}
then $\mc{U}\subseteq \mc{W} \subseteq \mc{V}^\star$. 
\end{cor}
\begin{pf}
For any $V \in \mc{W}$ and $V^\star \in \mc{V}^\star$,
\[\norm{V^\star - V}_\infty \leq \norm{V^\star - f_{C^k}(V^k)}_\infty + \norm{f_{C^k}(V^k) - V}_\infty \]
is satisfied for all $k \in \mathbb{N}$. Furthermore, by assumption, each $V\in\mc{W}$ has an associated iteration $V^{k+1} = f_{C^k}(V^k)$ whose limit point is equal to $V$, i.e. $\lim_{k\rightarrow\infty} \norm{f_{C^k}(V^k) - V}_\infty = 0 $. Additionally, 
\[\lim_{k \rightarrow\infty} \inf_{V^\star\in\mc{V}^\star} \norm{f_{C^k}(V^k) - V^\star}_\infty = 0,\]
follows from Proposition~\ref{prop:setConverge}. 
Therefore,
\[ \inf_{V^\star \in \mc{V}^\star} \norm{V^\star - V}_{\infty} \leq 0, \quad \forall \ V \in \mc{W}.\]
From the fact that the infimum over a compact set is always achieved for an element of the set~\cite{rudin1964principles}, $V = V^\star \in \mc{V}^\star$. Therefore $\mc{W} \subseteq \mc{V}^\star$. To see that $\mc{U} \subseteq \mc{W}$, take $C^k = C$ for all $k = 0, 1, \ldots$, then $\mc{U} \subseteq \mc{W}$. \qedwhite
\end{pf}
\begin{rem}
We make the distinction between $\mc{V}^\star$, $\mc{W}$, and $\mc{U}$ to emphasize that $\mc{V}^\star$ is not simply the set of fixed points corresponding to $f_C$ for all possible $C \in \mc{C}$, given by $\mc{U}$, or the limit points of $f_{C^k}$ for all possible sequences $\{C^k\}_{k \in \mathbb{N}} \subset \mc{C}$, given by $\mc{W}$. The fixed point set $\mc{V}^\star$ contains all possible limiting trajectories of $\{f_{C^k}(V^k)\}_{k\in\naturals}$ without assuming a limit point exists.
\end{rem}
In Corollary~\ref{cor:fCk_convergence}, $\mc{U}$ can be easily understood as the set of optimal value functions for the set of standard MDPs $([S], [A], P,C,\gamma)$ generated by $C \in \mc{C}$. An interpretation for $\mc{W}$ is perhaps less obvious. We use the following example to illustrate the differences between these three sets.

\begin{exmp}
Consider a single state, two action MDP with a discount factor $\gamma = 0.9$, where $\mc{C}$ is given by $\{\begin{bmatrix}0 & 1\end{bmatrix}, \begin{bmatrix}0 & 2\end{bmatrix}, \begin{bmatrix}1 & 1\end{bmatrix}\}$. Here, $\mc{U} = \{0, 10\}$ corresponds to the three optimal value functions when cost is fixed ---i.e., where $C^k = C \in \mc{C}$.  We note that if $\{C^k\} \subseteq \{\begin{bmatrix}0 & 1\end{bmatrix}, \begin{bmatrix}0 & 2\end{bmatrix}\}$, then $V^\star = 0$ regardless of how $C^k$ is chosen. Therefore $\mc{W} = \{0\} \cup \mc{U} = \mc{U}$. Finally, if $C^k$ is randomly chosen from $\mc{C}$ and $V^0 = 0$, $V^k$ will randomly fluctuate but satisfy $V^k \in \mc{V}^{\star} =  [0, 10]$.
\end{exmp}

In the context of robust MDPs, $\mc{U}$ contains all the fixed point value functions of regular MDPs. The value function set $\mc{W}$ contains the fixed point value functions that are \emph{invariant} to fluctuating costs within any subset of $\mc{C}$. On the other hand, if the value functions do not converge, the value function trajectory will still converge to $\mc{V}^\star$, even if $V^0 \notin \mc{V}^\star$. Therefore if the goal is to bound the asymptotic behaviour of $V^k$, it is more useful to determine $\mc{V}^\star$.

We summarize our results on set-based Bellman operator as the following: given a compact set of cost parameters $\mc{C}$, $F_{\mc{C}}$ converges to a unique compact set $\mc{V}^\star$. The set $\mc{V}^\star$ contains all the fixed points of $f_C$ for $C \in \mc{C}$. Furthermore, $\mc{V}^\star$ also contains the limit points of $f_{C^k}(V^k)$ for any $\{C^k\}_{k\in\naturals} \subseteq \mc{C}$, $V^0 \in \reals^{S}$, given that $\lim_{k\rightarrow \infty}V^k$ converges. Even if the limit does not exist, $V^k$ must asymptotically converge to $\mc{V}^\star$ in the Hausdorff distance. 
\section{Stochastic Games}\label{sec:connection2StochasticGames}
In this section, we further elaborate on the properties of the fixed point set $\mc{V}^\star$ in the context of stochastic games, and show that with an appropriate over-approximation of the Nash equilibria cost parameters, $\mc{V}^\star$ contains the optimal value functions for player one at Nash equilibria. 

A \emph{stochastic game} extends a standard MDP to a multi-agent competitive setting~\cite{shapley1953stochastic}. In the interest of clarity, we define Nash equilibria as well as player value functions in the context of two player stochastic games. However, the following definitions also extend to $N$ player stochastic games~\cite{filar2012competitive}.

We note that the stochastic games we discuss here implicitly assume \emph{imperfect information}~\cite[Def. 6.3.6]{filar2012competitive} --- at every state, both players have multiple actions to choose from. Therefore, each player's choice of action induces uncertainty in their opponent's costs.

In a two-player stochastic game, both players solve their own MDP while sharing the same states and dynamics. As opposed to standard MDPs, each player's cost and transition kernel depends on the \emph{joint policy}, $\pi = (\pi_1, \pi_2)$, where $\pi_1$ and $\pi_2$ are respectively player one and player two's policies as defined for standard MDPs in Section~\ref{sec:setup}. The set of joint policies is given by $\Pi$, while player one's and player two's sets of policies are given by $\Pi_1$ and $\Pi_2$, respectively. We denote the actions of player one by $a$ and the actions of player two by $b$. Players share a common state, given by $s \in [S]$. The transition kernel of the shared dynamics is determined by the tensor $Q \in \reals^{S\times S\times A_1 \times A_2}$, where $Q$ satisfies
\[\sum_{s'\in[S]} Q_{s'sab} = 1, \ \forall \ (s, a, b) \in [S]\times [A_1] \times [A_2],   \]
\[ Q_{s'sab}  \geq 0, \ \forall \ (s', s, a, b) \in [S]\times [S] \times [A_1] \times [A_2]. \]
Each player's cost is given by $D^i \in \reals^{S\times A_1\times A_2}$, where $D^1_{sab}$ and $D^2_{sab}$ denote player one and player two's cost when the joint action $(a,b)$ is taken from state $s$, respectively. 

For a specific policy adopted by player two, player one's transition kernel and cost can be represented using the same notation of Section~\ref{sec:setup}. When player two applies policy $\pi_2$, player one's transition kernel is given by
\begin{equation}\label{eqn:probabilityKernel}
P^1(\pi_2) \in \reals^{S\times SA_1}, \ P^1_{s',sa}(\pi_2)  =  \sum_{b \in [A_2]} (\pi_2)_{sb}Q_{s'sab}.    
\end{equation}
Furthermore, player one's cost is given by
\begin{equation}\label{eqn:stochasticGame_costDef}
C^1(\pi_2) \in \reals^{S\times A_1}, \  C^1_{sa}(\pi_2) = \sum_{b \in [A_2]} (\pi_2)_{sb} D^1_{sab}.    
\end{equation}
For a specific $\pi_1$ adopted by player one, player two's cost $C^2(\pi_1)$ and transition kernel $P^2(\pi_1) $ can be similarly defined. Each player then solves a discounted MDP given by $([S], [A_i], P^i(\pi_j), C^i(\pi_j), \gamma_i) $. Since each player only controls a part of the joint action space, the generalization to joint action space introduces \emph{non-stationarity} in the transition and cost, when viewed from the perspective of an individual player solving an MDP. 

Given a joint policy $(\pi_1, \pi_2)$, each player attempts to minimize its value function. Player $i$'s optimal discounted infinite horizon expected cost is given by 
\begin{equation}\label{eqn:expectedV_stochasticGame}
 V^i_{s} = \min_{\pi_i \in \Pi_i} \mathbb{E}^{\pi_i}_{s} \Big\{ \sum_{t = 0}^\infty \gamma_i^t C^i_{s^t a^t}(\pi_j)\Big\}, \quad \forall \ s \in [S].
\end{equation}

As formulated by~\eqref{eqn:expectedV_stochasticGame}, we denote the value function of player one, $V^1$, by $V \in \reals^S$ and the value function of player two, $V^2$, by $W \in \reals^S$. Given a joint policy $\pi$, both players have unique stationary value functions $\big(V(\pi_1, \pi_2), W(\pi_1, \pi_2)\big)$ given by 
\begin{subequations}\label{eqn:value_game}
\begin{align}
\label{eqn:value_game_player1}
    V(\pi_1, \pi_2)  = & \nu^1(\pi_1, \pi_2) + \gamma_1 M_{\pi_1}P^1(\pi_2)^\top V(\pi_1, \pi_2),\\
    W(\pi_1, \pi_2)  = & \nu^2(\pi_1, \pi_2) + \gamma_2 M_{\pi_2}P^2(\pi_1)^\top W(\pi_1, \pi_2),
\end{align}
\end{subequations}
where $\nu^1(\pi_1, \pi_2) = \sum_{i \in[S]} e_i e_i^\top M_{\pi_1}(\ones_s \otimes I_{A_1}) C^1(\pi_2)^\top e_i $ and $\nu^2(\pi_1, \pi_2) = \sum_{i \in[S]} e_i e_i^\top M_{\pi_2}(\ones_s \otimes I_{A_2}) C^2(\pi_1)^\top e_i $.
Since a stochastic game can be viewed as coupled MDPs, the MDP notion of optimality must be expanded to reflect dependency of a player's individual optimal policy on the joint policy space. We define a \emph{Nash equilibrium} in terms of each player's value function~\cite[Sec.3.1]{filar2012competitive}. 
\begin{defn}\label{def:NE}
[Two Player Nash Equilibrium]
A joint policy $\pi^\star = (\pi^\star_1, \pi^\star_2)$ is a Nash equilibrium if the corresponding value functions as given by~\eqref{eqn:value_game} satisfy
\[ V(\pi_1^\star, \pi_2^\star) \leq V(\pi_1, \pi_2^\star), \quad \forall \ \pi_1 \in \Pi_1,\]
\[ W(\pi_1^\star, \pi_2^\star) \leq W(\pi_1^\star, \pi_2), \quad \forall \ \pi_2 \in \Pi_2. \]
We denote the Nash equilibrium value functions as $V^\star = V(\pi_1^\star, \pi_2^\star)$, $W^\star = W(\pi_1^\star, \pi_2^\star)$ and the set of Nash equilibria for a stochastic game as $\Pi_{NE} \subset \Pi$.
\end{defn}
Definition~\ref{def:NE} implies that a Nash equilibrium is achieved when the joint policy simultaneously generates both value functions $V^\star$ and $W^\star$, which are the fixed points of the Bellman operator with respect to parameters  $\Big(C^1(\pi_2), P^1(\pi_2)\Big)$ and $\Big(C^2(\pi_1), P^2(\pi_1)\Big) $, respectively --- i.e.
$V^\star = \underset{\pi_1 \in \Pi_1}{\min} \Big\{\nu^1(\pi_1, \pi^\star_2) + \gamma_1 M_{\pi_1}P^1(\pi^\star_2)^\top V^\star\Big\}$, and  
$W^\star = \underset{\pi_2 \in \Pi_2}{\min} \Big\{\nu^2(\pi^\star_1, \pi_2) + \gamma_2 M_{\pi_2}P^2(\pi^\star_1)^\top W^\star\Big\}$.

A Nash equilibrium is not unique for general sum stochastic games. Furthermore, Nash equilibria policies are not necessarily composed of deterministic individual policies. Therefore while each player's Nash equilibrium value function is always the fixed point of the associated Bellman operator, the Nash equilibrium policy for each player is \emph{not} the optimal deterministic policy associated to the Nash equilibrium value function in general. The existence of at least one Nash equilibrium for any general sum stochastic game is given in~\cite{filar2012competitive}. When the stochastic game is also zero sum, all Nash equilibria correspond to a unique value function. 

Since the technical content of this paper does not address non-stationarity in the transition kernel, we focus on analyzing non-stationarity in the cost term.  Specifically, we constrain our analysis to a \emph{single controller game}~\cite{filar2012competitive}, i.e. when the transition kernel is controlled by player one only. Single controller stochastic games form an important class of games that models dynamic control in queueing networks~\cite{altman1994flow} and attacker-defender games with stochastic transitions~\cite{ang2017game,eldosouky2016single}. Similar to our discussion of a two player Nash equilibrium, we exclusively consider a two player single controller game. However, we note that the following definition can be extended to an $N$ player single controller stochastic game in which the transition kernel is independent of all but one player's actions.

\begin{defn}[Single controller game]\label{def:singleControllerGame}
A single controller game is a two player stochastic game where the probability transition kernel is independent of player two's actions, i.e., for each $(s', s, a) \in [S]\times [S]\times[A]$
\[
Q_{s'sab} = Q_{s'sab'}, \quad  \forall \ b, b' \in [A],
\]
i.e. $P^1(\pi_2) = P$, $\forall \ \pi_2 \in \Pi_2$ and $P^2(\pi_1)_{s',sb} = P^2(\pi_1)_{s',sb'}$, $\forall \ b, b' \in [A], \ \pi_1 \in \Pi_1$.
\end{defn}

Although both players are still optimizing their value functions in a single controller game, player two's policy only affects its immediate cost at each state, while its transition dynamic becomes a time-varying Markov chain. However, player two's policy still affects player one's MDP through cost matrix $C^1(\pi_2)$.

We analyze a single controller game from the set-based MDP perspective by utilizing Proposition~\ref{prop:setConverge}. Suppose we are given a compact set $\mc{C} \subset \reals^{S\times A}$ that over-approximates the set of $\mc{C}^{NE} $ --- i.e. cost parameters that player one observes at Nash equilibria, 
\begin{equation}\label{eqn:overApproxCost_stochasticGame}
    \mc{C}^{NE} = \{C^1(\pi^\star_2) \in \reals^{S\times A}\ | \ (\pi^\star_1, \pi^\star_2) \in \Pi_{NE} \} \subseteq \mc{C}.  
\end{equation}
Then we show that the Nash equilibria value functions belong to the fixed point set of $F_{\mc{C}}$. 

Valid over-approximations to $\mc{C}^{NE}$ can be easily found --- the simplest being the interval set of all feasible costs.
\begin{exmp}[Interval set approximation]\label{ex:intervalSets}
An approximation to  $\mc{C}^{NE}$ can always be given by interval sets. At each state-action pair $(s,a)$, the MDP cost parameter for player one is given by~\eqref{eqn:stochasticGame_costDef}. Then we can take the maximum and minimum elements of the set $\{D^1_{sab}\}_{b \in [A_2]}$ for all state actions pairs $(s, a)$ to form an interval set $\mc{C} = \mc{C}_{11}\times\ldots\times\mc{C}_{SA_1} \in H(\reals)^{S\times A_1}$, such that
\[\mc{C}_{sa} = \{D^1_{sab}\}_{b \in [A_2]} = [\underline{C}_{sa}, \overline{C}_{sa}],\]
where $\underline{C}_{sa} = \min_{b \in [A_2]} D^1_{sab}$ and $\overline{C}_{sa} = \max_{b \in [A_2]} D^1_{sab}$ can be directly observed.

\end{exmp}

Interval sets will always give an admissible approximation. However, more general sets such as polytopes allow for more precise representations of the limiting value function trajectories for the game player.  

\begin{figure}[ht]
    \centering
    \includegraphics[width = 0.6\linewidth]{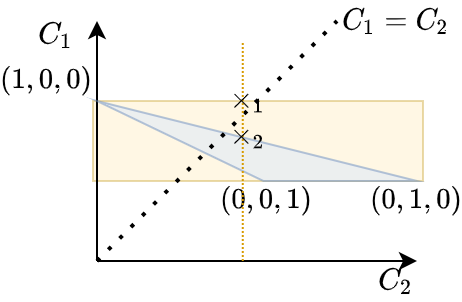}
    \caption{Feasible player costs vs interval set over-approximation.}
    \label{fig:NE_polytope}
\end{figure}

\begin{exmp}[Polytope set approximation]\label{ex:polytopeSets}
Consider the set of costs at a particular state $s$ in a two player single controller stochastic game, for which $A_1 = 2$ and $A_2 = 3$. Player one's costs corresponding to player two's deterministic policies are given by points $(1,0,0)$, $(0,1,0)$, and $(0,0,1)$ in Figure~\ref{fig:NE_polytope}. Any mixed policy from player two will result in an expected cost for player one that corresponds to a point within the blue region in Figure~\ref{fig:NE_polytope}. On the other hand, the interval set approximation from Example~\ref{ex:intervalSets} is given by the yellow polytope. In this example, we can observe that the interval set is a generous over-approximation of player one's feasible costs.

An over-approximation of the set of feasible costs also over-approximates possible limiting trajectories for a player's learning algorithm. We consider the point $\times_1 = (C'_2, C'_1)$ in Figure~\ref{fig:NE_polytope}. Fixed at this cost, value iteration would choose $a_2$ corresponding to $C'_2$, and return the corresponding discounted value function and transition kernel from state $s$. However, the \emph{feasible cost} when action $a_2$ has equivalent cost $C'_2$ is at $\times_2 = (\bar C_1, C'_2)$ on the boundary of the blue polytope. Since $\times_2$ lies below the line $C_1 = C_2$, $a_1$ corresponding to $\bar C_1$ is actually the optimal action. Therefore, the resulting cost and transition kernel would have been different. This corresponds to a different value function trajectory that would have been infeasible.
\end{exmp}

The set of feasible costs itself is an over-approximation of the set of Nash equilibria costs $\mc{C}^{NE}$. As Example~\ref{ex:polytopeSets} shows, the extension from interval sets to compact sets enables additional information (feasible costs, knowledge of opponents' action constraints) to be used to approximate $\mc{C}^{NE}$ to greater accuracy. 

Given a compact set $\mc{C}$ that over-approximates the set of player one's cost parameters at Nash equilibria, $\mc{C}^{NE}$, we now show that the Nash equilibria value functions for player one must lie within $\mc{V}^\star$, the fixed point set of $F_{\mc{C}}$. 

\begin{thm}
In a single controller game, let $\mc{C}\subset \reals^{S\times A}$ be an over-approximation of Nash equilibria costs for player one as in~\eqref{eqn:overApproxCost_stochasticGame}. 
If $\mc{C}$ is compact, then the set of stationary value functions for player one at Nash equilibria policies~\eqref{eqn:value_game_player1} is a subset of $\mc{V^\star}$, the fixed point set of $F_{\mc{C}}$.
\end{thm}
\begin{pf}
We define the set of Nash equilibria value functions for player one as 
\begin{equation}
    \begin{aligned}\label{eqn:valueSet_NE}
        \mc{V}^{NE} = \Big\{ &V \in \reals^S \  | \ V = f_{C^1(\pi^\star_2)}(V)\Big\},
    \end{aligned}
\end{equation}
where the Bellman operator $f_{C^1(\pi^\star_2)}$ is defined with $P$, the $\pi_2$-independent transition kernel for both players. 
For any $V^\star \in \mc{V}^{NE}$, there exists $C^\star = C^1(\pi^\star_2) \in \mc{C}$ such that $V^\star$ is the fixed point of $f_{C^\star}$. Then from Corollary~\ref{cor:fCk_convergence}, $V^\star \in \mc{V}^\star$. 
\qedwhite
\end{pf}
\begin{rem}
Although the Nash equilibrium value function $V^\star$ is always the unique fixed point of $f_{C^\star}$ given by~\eqref{eqn:bellmanOp}, where $C^\star$ is player one's cost at Nash equilibrium, we note that in general, player one's policy at Nash equilibrium is not the optimal deterministic policy of $f_{C^\star}(V^\star)$ given by~\eqref{eqn:optimalPol}; this is because the joint policy at Nash equilibrium may not be composed of deterministic individual policies, while the solution to~\eqref{eqn:optimalPol} is always deterministic. 

However, if we consider the set of all deterministic policies that solves~\eqref{eqn:optimalPol}, then player one's policy at Nash equilibrium must be a convex combination within this set
~\cite{filar2012competitive}. 
\end{rem}

We summarize the application of set-based MDP framework to single controller stochastic games as the following: when $\mc{C}$ over-approximates the set of costs at Nash equilibria, the fixed point set $\mc{V}^\star$ of operator $F_{\mc{C}}$  contains all of the Nash equilibria value functions for player one in a single controller stochastic game. 

\section{Application to Interval Set-Based Bellman Operator}
\label{sec:ex}
In this section, we show that when the cost parameter set $\mc{C}$ and the initial value function set $\mc{V}^0$ are interval sets, the fixed point set $\mc{V}^\star$ of $F_{\mc{C}}$ is also an interval set, as done similarly in~\cite{givan2000bounded}. However, we note that convergence in~\cite{givan2000bounded} is shown under an unconventional partial ordering scheme. Leveraging our set-based Bellman operator framework and the Hausdorff distance as our metric, our result is derived in a much more straightforward manner using interval arithmetic. 

As shown in Example~\ref{ex:intervalSets} and Example~\ref{ex:polytopeSets}, one over-approximation of the set of Nash equilibria costs is given by interval sets. In this section we show how to compute the fixed point set $\mc{V}^\star$ of an interval set-based Bellman operator. Suppose the set $\mc{C}$ is given by 
\begin{equation}\label{eqn:intervalC}
\mc{C} = \Big\{C \in \reals^{S\times A} \ | \ C_{sa} \in [\underline{C}_{sa}, \overline{C}_{sa}], \ \forall\ (s,a) \in [S]\times[A] \Big\}.
\end{equation}
and the input value function set is given by 
\begin{equation}\label{eqn:invervalV}
\mc{V} = \Big\{V \in \reals^{S} \ | \ V_s \in [\underline{V}_{s}, \overline{V}_{s}], \ \forall\ s \in [S]\Big\}.
\end{equation}
\subsection{Hausdorff distance between interval sets}
We first show that the Hausdorff distance between two interval sets $\mc{V}, \mc{V}' \in \ H(\reals^S)$ can be computed by only comparing the upper and lower bounds of the intervals themselves. 
\begin{lem}\label{lem:intervalHausdorff}
For two intervals $\mc{X}, \mc{Y} \in H(\reals^S, \norm{\cdot}_\infty)$ given by $\mc{X} = \intervalSet{x}$, $\mc{Y} = \intervalSet{y}$, where $\underline{x}, \overline{x}, \underline{y}, \overline{y} \in \reals^S$, the Hausdorff distance between $\mc{X}$ and $\mc{Y}$ can be calculated as 
\[d_H(\mc{X},\mc{Y}) = \max\{\norm{\underline{x} - \underline{y}}_\infty, \norm{\overline{x} - \overline{y}}_\infty\}.\]
\end{lem}
\begin{pf}
We consider the component-wise Hausdorff distance by noting that when coupled with the infinity norm on $\reals^S$, the Hausdorff distance satisfies
\[d_H(\mc{X}, \mc{Y}) = \max_{i \in [S]} \ d_H(\mc{X}_i, \mc{Y}_i),\]
where $\mc{X} = \mc{X}_1\times\ldots\mc{X}_S$ and $\mc{Y} = \mc{Y}_1\times\ldots\mc{Y}_S$~\cite{chavent2004hausdorff}. 

We first compute $d_H(\mc{X}_i, \mc{Y}_i)$, where $\mc{X}_i  = [\underline{x_i}, \overline{x_i}]$ and $\mc{Y}_i  = [\underline{y_i}, \overline{y_i}]$ are interval sets. Recall that the infinity norm can be written using $\max$ operators given in~\eqref{eqn:infNorm2Max}. The nested $\max$ representation of the infinity norm allows 
us to directly evaluate the infimum and supremum of $\norm{x_i - y_i}_\infty$ over $\mc{X}_i$ and $\mc{Y}_i$ respectively, as 
\[\sup_{y_i \in \mc{Y}_i} \inf_{x_i \in \mc{X}_i} \norm{x_i - y_i}_\infty = \max\{\max(\underline{x}_i - \underline{y}_i), \max(\overline{y}_i - \overline{x}_i)\}.\]
Similarly, we can derive
\[\sup_{x_i \in \mc{X}_i} \inf_{y_i \in \mc{Y}_i} \norm{x_i - y_i}_\infty = \max\{\max(\underline{y}_i - \underline{x}_i), \max(\overline{x}_i - \overline{y}_i)\}.\]
Finally we recall the definition of Hausdorff distance: 
\begin{equation}
    \begin{aligned}
        d_H(\mc{X}_i, \mc{Y}_i) & = \max \{ \sup_{y_i \in \mc{Y}_i} \inf_{x_i \in \mc{X}_i} \norm{x_i - y_i}_\infty, \\
        & \qquad \quad \ \sup_{x_i \in \mc{X}_i} \inf_{y_i \in \mc{Y}_i} \norm{x_i - y_i}_\infty \} \\
        & = \max\{\max(\underline{x_i} - \underline{y_i}), \max(\overline{y_i} - \overline{x_i}), \\
        & \qquad \quad \ \max(\underline{y_i} - \underline{x_i}), \max(\overline{x_i} - \overline{y_i}) \} \\
        & = \max\{ \norm{ \underline{x_i} - \underline{y_i} }_\infty, \norm{\overline{x_i}- \overline{y_i}}_\infty\}.
    \end{aligned}
\end{equation}
Then the total Hausdorff distance between $\mc{X}$ and $\mc{Y}$ is given by 
\begin{equation}
    \begin{aligned}
        d_H(\mc{X}, \mc{Y}) & = \underset{i \in [S]}{\max} \{ \max\{\norm{\underline{x_i} - \underline{y_i}}_\infty, \norm{\overline{x_i} - \overline{y_i}}_\infty\} \} \\
        & = \max\{\norm{\underline{x} - \underline{y}}_\infty, \norm{\overline{x} - \overline{y}}_\infty\}.
    \end{aligned}
\end{equation} \qedwhite
\end{pf}
Lemma~\ref{lem:intervalHausdorff} shows that interval sets are \emph{nice} in that their Hausdorff distances can be derived via component-wise operations on the boundaries of the intervals. We use Lemma~\ref{lem:intervalHausdorff} later in this section to obtain convergence guarantees of set-based value iteration to the fixed point set of the interval set-based Bellman operator. 
\subsection{Interval arithmetic}
To compute the fixed point of an interval set-based Bellman operator, we introduce some relevant interval arithmetic operators~\cite{moore1966interval}. 
\begin{equation}\label{eqn:intervalArith}
    \begin{aligned}
        \alpha[a,b] & = [\alpha a, \alpha b], \quad \alpha \geq 0 ,\\
        [a, b] + [c, d] & = [a + c, b + d], \\
        [a, b] - [c, d] & = [a - d, b - c], \\
        \min\{[a, b], [c, d]\} &  = [\min\{a, c\}, \min\{b, d\}],
    \end{aligned}
\end{equation}
where the last operator $\min\{[a,b], [c,d]\}$ denotes the smallest interval that contains $\{ \min\{x,y\}, | \ x \in [a,b], y \in [c,d]\}$.
The equivalence relationship for the $\min$ operator in~\eqref{eqn:intervalArith} is not as obvious as the standard addition and subtraction operators. Here we prove that the equivalence for $\min$ operator is indeed as given. 

\begin{lem}\label{lem:minSetOperator}
The $\min$ operator for interval sets can be calculated as 
\[ \min\{[a, b], [c, d]\}   = [\min\{a, c\}, \min\{b, d\}].
\]
\end{lem}
\begin{pf}
Let us consider the sets $\mc{A} = \min\{[a, b], [c, d]\}$ and $\mc{B} = [ \min\{a, c\},  \min\{ b, d\} ]$. We first show that $\mc{A} \subseteq \mc{B}$: for $z \in \min\{[a, b], [c, d]\}$, there exists $x \in [a,b]$ and $y \in [c,d]$ such that $z = \min\{x,y\}$. Then necessarily, $\min\{a,c\} \leq z$ and $z \leq \min\{b,d\}$ must be satisfied. 

To prove the inclusion $\mc{B} \subseteq \mc{A}$, take $v \in [\min\{a, c\}, \min\{b, d\}]$.  If $v \in [a,b]$, then $v  = \min \{ v, \max \{ v, d \} \}$. 
If $\max\{v,d\} = d$, then $v \in\min\{[a, b], [c, d] \}$ follows from $v \in [a,b]$ and $d \in [c,d]$. If $\max\{v,d\} = v$, then $d < v \leq b$. This contradicts $v \in [\min\{a, c\}, \min\{b, d\}]$. 

If $v \notin [a,b]$, then either $a \neq \min\{a,c\}$  or $b \neq \min\{b,d\}$. This is equivalent to either $c \leq v < a$ or $b < v \leq d$ is true. $b < v \leq d$ cannot be true since $v \in [\min\{a, c\}, \min\{b, d\}]$. 
$c \leq v < a$ implies that $v \in [c,d]$ and $v = \min\{v, a\}$, then $v \in \min\{[a, b], [c, d] \}$. \qedwhite
\end{pf}
With Lemma~\ref{lem:intervalHausdorff} and~\ref{lem:minSetOperator}, we can analytically compute the fixed point set of an interval set-based Bellman operator and give convergence guarantees of interval set-based value iteration.  
\begin{prop}\label{prop:intervalVIBounds}
For interval sets $\mc{C}$ and $\mc{V}$ given by~\eqref{eqn:intervalC} and~\eqref{eqn:invervalV}, respectively, $F_{\mc{C}}(\mc{V})$ defined in Definition~\ref{def:setBasedBellman} is an interval set that satisfies
\[F_{\mc{C}}(\mc{V}) = \{V \ | \ V \leq V_u, \ -V \leq -V_l, \ V \in \reals^{S}\},\] 
for $V_l = f_{\underline{C}}(\underline{V})$ and $V_u = f_{\overline{C}}(\overline{V})$. 

Furthermore, the sequence $\{\mc{V}^k\}_{k \in \mathbb{N}}$ generated by the iteration $\mc{V}^{k+1} = F_{\mc{C}}(\mc{V}^k)$ starting from any interval set $\mc{V}^0 $ will converge to $\mc{V}^\star$ in Hausdorff distance: for every $\epsilon > 0$, there exists $\mc{V}^k$ which satisfies \begin{equation}\label{eqn:setbasedConvergence}
d_H(\mc{V}^k, \mc{V}^\star) \leq \epsilon/2,
\end{equation}
where~\eqref{eqn:setbasedConvergence} is satisfied if $d_H(\mc{V}^{k}, \mc{V}^{k-1})\frac{2\gamma}{1 - \gamma} < \epsilon$. 
\end{prop}
\begin{pf}
We recall the set-based Bellman operator Definition~\ref{def:setBasedBellman} and the component-wise definition of $f_C$ in Definition~\ref{def:bellmanOp}. Using these definitions and the fact that $\mc{C} = [\underline{C}, \ \overline{C}]$ and $\mc{V} = [\underline{V}, \overline{V}]$ are both interval sets, the set-based Bellman operator can be written as
\[\Big(F_{\mc{C}}(\mc{V})\Big)_s = \cl \underset{\substack{C \in \intervalSet{C} \\ V \in \intervalSet{V}}}{\bigcup}  \min_{a \in[A]} C_{sa} + \gamma \sum_{s' \in [S]} P_{s',sa}V_{s'}.\] 
Let $G(C_{sa}, V) = C_{sa} + \gamma\sum_{s' \in [S]} P_{s',sa}V_{s'} $. $G$  is a continuous function and order preserving in its inputs $C_{sa}$ and $V$. Therefore the union over interval sets in $\big(F_{\mc{C}}(\mc{V})\big)_s$ can be written using interval arithmetic notation as
\begin{align}\label{eqn:intervalSetPf1}
    & \underset{\substack{C \in \intervalSet{C} \\ V \in \intervalSet{V}}}{\bigcup}  \min_{a \in[A]} C_{sa} + \gamma \sum_{s' \in [S]} P_{s',sa}V_{s'} \\
    = & \{ \min_{a \in[A]} C_{sa} + \gamma \sum_{s' \in [S]} P_{s',sa}V_{s'} \ | \ C \in \intervalSet{C},  V \in \intervalSet{V}\} \\
    = & \min_{a\in[A]} \intervalSet[sa]{C} + \gamma \sum_{s'\in[S]}P_{s',sa}\intervalSet[s']{V}.\label{eqn:intervalSetPf2}
\end{align}
Since interval sets are closed by definition, the closure of~\eqref{eqn:intervalSetPf1} must also equal~\eqref{eqn:intervalSetPf2}. Therefore, $F_{\mc{C}}(\mc{V})$ can be equivalently written component-wise as 
\begin{equation}\label{eqn:intervalSetPf3}
\Big(F_{\mc{C}}(\mc{V})\Big)_s = \ \min_{a \in[A]}\ \intervalSet[sa]{C} + \gamma \sum_{s' \in [S]} P_{s',sa}\intervalSet[s']{V}.
\end{equation}
Then, $\gamma > 0$ and $P_{s',sa} \geq 0$ for all $ (s', s, a) \in [S]\times[S]\times[A]$ allow us to directly perform interval arithmetic component-wise for $F_{\mc{C}}$ as
\begin{subequations}
    \begin{align}
         \Big(F_{\mc{C}}(\mc{V})\Big)_s & = \min_{a \in [A]} \Big[\underline{C}_{sa} + \notag \gamma\sum_{s'\in[S]}P_{s',sa}\underline{V}_{s'},\\ 
         & \quad \quad \quad \quad \overline{C}_{sa} + \gamma \sum_{s'\in[S]}P_{s',sa}\overline{V}_{s'}\Big] \label{subeqn:intervalOps_b}\\
         & = \Big[\min_{a \in [A]}\underline{C}_{sa} + \gamma\sum_{s'\in[S]}P_{s',sa}\underline{V}_{s'}, \notag\\
         &\quad \quad \min_{a \in [A]}\overline{C}_{sa} + \gamma\sum_{s'\in[S]}P_{s',sa}\overline{V}_{s'} \Big] \label{subeqn:intervalOps_d}\\
         & = [\Big(f_{\underline{C}}(\underline{V})\Big)_s, \ \Big(f_{\overline{C}}(\overline{V})\Big)_s], \label{subeqn:intervalOps_e}
    \end{align}
\end{subequations}
where~\eqref{subeqn:intervalOps_d} utilizes the interval set-based minimization derived in Lemma~\ref{lem:minSetOperator}, and~\eqref{subeqn:intervalOps_e} follows from Definition~\ref{def:bellmanOp}. 

The image of $F_{\mc{C}}$ is another closed interval, as shown by~\eqref{subeqn:intervalOps_e}. From Theorem~\ref{thm:uniqueSetFixedPont}, any interval set $\mc{V}^0  = \intervalSet{V}$ generates an iteration $\mc{V}^{k+1} = F_{\mc{C}}(\mc{V}^k)$ which satisfies 
$\lim_{k \rightarrow \infty} F_{\mc{C}} (\mc{V}^k) = \mc{V}^\star$. We can use interval arithmetic to derive $\mc{V}^\star = \lim_{k\rightarrow \infty} F_{\mc{C}}(\mc{V}^k) = \Big[  \lim_{k\rightarrow \infty}f_{\underline{C}}(\underline{V^k}),  \lim_{k\rightarrow \infty}f_{\overline{C}}(\overline{V^k}) \Big] = \intervalSet{V^\star}$,  
where $\underline{V^\star}$ and $\overline{V^\star}$ are the fixed points of $f_{\underline{C}}$ and $f_{\overline{C}}$, respectively. 

At each iteration, the Hausdorff distance between $\mc{V}^k$ and $\mc{V}^\star$ is given by $d_{H}(\mc{V}^k, \mc{V}^\star) = d_H\Big([f_{\underline{C}}(\underline{V^k}), f_{\overline{C}}(\overline{V^k})], \intervalSet{V^\star}\Big)$.
Using Lemma~\ref{lem:intervalHausdorff}, $ d_{H}(\mc{V}^k, \mc{V}^\star)$ is given by \[ \max \Big\{\norm{f_{\underline{C}}(\underline{V^k}) - \underline{V^\star}}_\infty , \norm{f_{\overline{C}}(\overline{V^k}) - \overline{V^\star}}_\infty \Big\} .\] Similarly, $d_H(\mc{V}^{k+1}, \mc{V}^k)$ is given by
\begin{align*}
\max\Big\{ &  \norm{f_{\underline{C}}(\underline{V^k}) - f_{\underline{C}}(\underline{V^{k+1}})}_\infty, \\
 & \norm{f_{\overline{C}}(\overline{V^k}) - f_{\overline{C}}(\overline{V^{k+1}})}_\infty \Big\}.
\end{align*}
From Lemma~\ref{thm:stoppingCriteria}, if $\norm{f_{\underline{C}}(\underline{V^k}) -  f_{\underline{C}}(\underline{V^{k+1}})}_\infty  < \epsilon \frac{1 - \gamma}{2\gamma} $ for some $\epsilon > 0$, then $ \norm{f_{\underline{C}}(\underline{V^k}) - \underline{V^\star}}_\infty < \frac{\epsilon}{2}$. 
Similarly, if $\norm{f_{\overline{C}}(\overline{V^k}) -  f_{\overline{C}}(\overline{V^{k+1}})}_\infty  < \epsilon \frac{1 - \gamma}{2\gamma} $ for some $\epsilon > 0$, then $ \norm{f_{\overline{C}}(\overline{V^k}) - \overline{V^\star}}_\infty < \frac{\epsilon}{2}$. Therefore if $\max\Big\{\norm{f_{\underline{C}}(\underline{V^k}) - \underline{V^\star}}_\infty, \norm{f_{\overline{C}}(\overline{V^k}) -  f_{\overline{C}}(\overline{V^{k+1}})}_\infty \Big\} < \epsilon$, then $\max\Big\{\norm{f_{\underline{C}}(\underline{V^k}) - \underline{V^\star}}_\infty, \norm{f_{\overline{C}}(\overline{V^k}) - \overline{V^\star}}_\infty\Big\} <  \frac{\epsilon}{2}$. Since  $d_H(\mc{V}^{k+1}, \mc{V}^k) = \max\Big\{\norm{f_{\underline{C}}(\underline{V^k}) - \underline{V^\star}}_\infty$, $\norm{f_{\overline{C}}(\overline{V^k}) -  f_{\overline{C}}(\overline{V^{k+1}})}_\infty \Big\}$
and $d_H(\mc{V}^{k+1} - \mc{V}^\star)  = \max\Big\{\norm{f_{\underline{C}}(\underline{V^k}) - \underline{V^\star}}_\infty,$ $\norm{f_{\overline{C}}(\overline{V^k}) - \overline{V^\star}}_\infty\Big\}$, we conclude that if $d_H(\mc{V}^{k+1}, \mc{V}^k) < \frac{(1 - \gamma)\epsilon}{2\gamma}$, then $d_H(\mc{V}^{k+1} - \mc{V}^\star) < \frac{\epsilon}{2}$.\qedwhite
\end{pf}
\begin{rem}
In existing work, $V_{l}$ is equivalent to the \emph{optimistic value function} in~\cite{iyengar2005robust} when the transition kernel is known and cost uncertainty is given by bounded intervals. Furthermore, 
Proposition~\ref{prop:intervalVIBounds} specializes interval value iteration from~\cite{givan2000bounded} to cost uncertainty only and proves stronger convergence results due to this specialization.
\end{rem}

We note that while $\mc{V}^k$ converges to $\mc{V}^\star$ in Hausdorff distance, it is not an over approximation to $\mc{V}^\star$. In fact, if $\underline{V}^k > \underline{V^\star}$ for some $k \in \mathbb{N}$, then each $\mc{V}^\star \subsetneq \mc{V}^k $ for all $k$. Nonetheless, we can still utilize $\mc{V}^k$ to obtain an \emph{over-approximation} of $\mc{V}^\star$ by using estimate intervals $\tilde{\mc{V}}^{k+1} = [f_{\underline{C}}(\underline{V^k}) - \ones_S\epsilon, f_{\underline{C}}(\underline{V^k})+ \ones_S\epsilon]$.

\section{Numerical Example}
\label{sec:stochasticGames}
Our analysis of the set-based Bellman operator is motivated by dynamic programming-based learning algorithms in stochastic games. We demonstrate this by applying interval set-based value iteration in a two player single controller stochastic game, and showing that both transient and asymptotic behaviours of player one's value function can be bounded, regardless of the opponent's learning algorithm. 

We consider a two player single controller stochastic game as defined in Definition~\ref{def:singleControllerGame}, where each player solves a discounted MDP given by $([S],[A_{1,2}], P, C^{1,2}, \gamma_{1,2}) $, where $A_1 = A_2 = A$. Both players share an identical state-action space $([S],[A])$ as well as the same transition probabilities $P$ controlled by player one's actions. Player one's cost is given by
\[C^1_{sa}(\pi_2) = C_{sa} + J_{sb}\pi_2(s,b),  \quad \forall \ (s,a) \in [S]\times[A],\]
while player two's cost is given by 
\[C^2_{sb}(\pi_1) = C_{sb} - J_{sa}\pi_1(s,a),  \quad \forall \ (s,b) \in [S]\times[A],\]
where the matrix $J \in \reals_+^{S\times A}$ is the same for the two costs. 

While algorithms that converge to Nash equilibrium exist~\cite{littman1994markov,hu2003nash} for such single controller games, convergence is not guaranteed if players do not coordinate on which algorithm to use between themselves.
In this section, we utilize the set-based Bellman operator to show that we can determine the value function set that player one's Nash equilibrium value function belongs to, and equivalently, the value function set that player one's value function trajectory converges to, regardless of what the opponent does.

\begin{figure}[ht]
\centering
\begin{subfigure}{.45\columnwidth}
  \centering
  \includegraphics[width=0.9\linewidth]{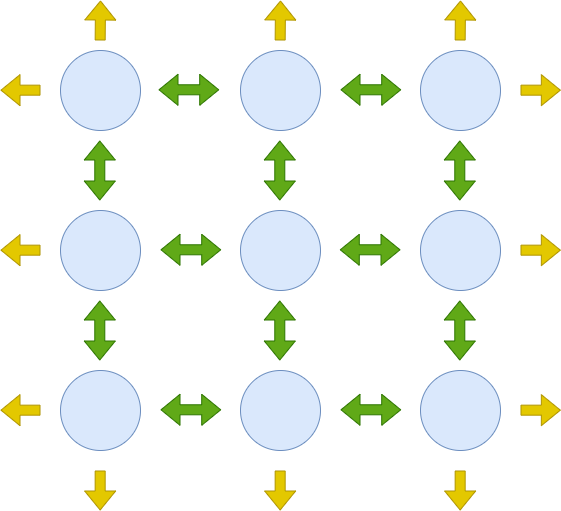}
  \caption{}
  \label{fig:grid_MDP}
\end{subfigure}%
\begin{subfigure}{.45\columnwidth}
  \centering
  \includegraphics[width=0.9\linewidth]{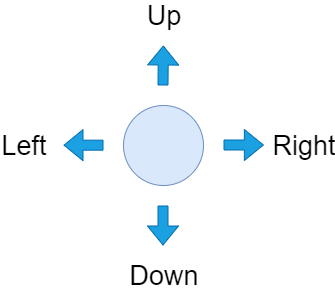}
  \caption{}
  \label{fig:action_MDP}
\end{subfigure}
\caption{(a): Each player's state space $[S]$, $S = 9$. Green actions leads to a neighbouring state and yellow actions are infeasible. (b): Actions space $[A]$, $A = 4$.}
\label{fig:test}
\end{figure}

We define the state space of a stochastic game on a $3\times 3$ grid, shown in Figure~\ref{fig:grid_MDP}, where the total number of states is $S = 9$ and total number of actions per state is $A = 4$. State $s'$ is a neighbouring state of $s$ if it is immediately connected to $s$ by a green arrow in Figure~\ref{fig:grid_MDP}. At state $s$, let $\mc{N}_s$ denote the set containing all neighbouring states of $s$ and let $N_s$ denote the number of elements in $\mc{N}_s$. 

As shown in Figure~\ref{fig:action_MDP}, the actions available in each state are labelled `left', `right', `up', or `down'. From each state $s$, an action is feasible if it coincides with a green arrow in Figure~\ref{fig:grid_MDP}, and infeasible if it coincides with a yellow arrow. For feasible actions, its transition probabilities are given as
\begin{equation}\label{eqn:feasible_action_transitions}
    P_{s'sa} = \begin{cases}
    0.7 & s' = \text{ target state} \\
    \frac{0.3}{N_s-1} & s' \neq\text{ target state}, \ s' \in \mc{N}_s\\ 
    0 & \text{otherwise}
    \end{cases}.
\end{equation}
In~\eqref{eqn:feasible_action_transitions}, we define the target state $s'$ of state-action pair $(s,a)$ to be the neighbouring state of $s$ in the direction of action $a$.
If action $a$ is infeasible, its transition probabilities are defined as 
\begin{equation}
    P_{s'sa} = \begin{cases}
    \frac{1}{N_s} &  s' \in \mc{N}_s\\ 
    0 & \text{otherwise}
    \end{cases}.
\end{equation}
We select matrices $C, J \in \reals^{9\times 4}$ by randomly sampling each element $C_{sa}, J_{sa}$ uniformly from the interval $[0,1]$. As in Example~\ref{ex:intervalSets}, we derive an over-approximation of player one's feasible costs as interval set $\mc{C}$, given by
\begin{equation}
\Big\{C^1 \in \reals^{9\times 4} \ | \ C^1_{sa} \in [{C}_{sa}, {C}_{sa} + J_{sa}], \ \forall\ (s,a) \in [9]\times[4] \Big\},
\end{equation}
where the upper bound $C_{sa} + J_{sa}$ is achieved when player two's probability of taking action $b = a$ from state $s$ is $1$. 

We consider a two player value iteration algorithm presented in Algorithm~\ref{alg:VI_stochasticGames} which forms the basis of many dynamic programming-based learning algorithms for stochastic games~\cite{filar2012competitive,littman1994markov}. At each time step, player one takes the optimal policy $\pi^{k+1}$ given by~\eqref{eqn:optimalPol} that solves the Bellman operator $f_{C^k}(V^k)$, where $C^k$ is player one's cost parameter at step $k$ and $V^k$ is player one's value function at step $k$ --- i.e. player one performs value iteration at every time step. Player two obtains its optimal policy using function $g: \Pi_1 \rightarrow \Pi_2$, we do not make any assumptions of $g$, it may produce any policy $\pi_2$ in response to the policy $\pi_1$. 
\begin{algorithm}
\caption{Two player VI}
\begin{algorithmic}[h]
\Require \(([S],[A],P,C^{1,2}, \gamma_{1,2})\), \(V_0\).
\Ensure \( V^\star, \pi^\star_1\)
\State{\(\pi_1^0(s) =\pi_2^0(s) = 0\), \(\forall \ s \in [S]\)}
\For{\(k = 0, \ldots, \)}
\State{\(C = C^1(\pi^k_1, \pi^k_2)\)}
\State{\((V^{k+1}, \pi_1^{k+1}) = f_{C}(V^k)\)}
\State{\(\pi^{k+1}_2 = g(\pi^{k+1}_1) \)}
\EndFor 
\end{algorithmic}
\label{alg:VI_stochasticGames}
\end{algorithm}

Our analysis provides bounds on player one's value function when we do not know how player two is updating its policy --- i.e. when $g$ is unknown. In simulation, we take $g$ to be different strategies and show that player one's value functions are bounded by the interval set analysis and converges towards the fixed point set of the corresponding Bellman operator.

\begin{figure}[ht]
    \centering
    \includegraphics[width=1.04\columnwidth]{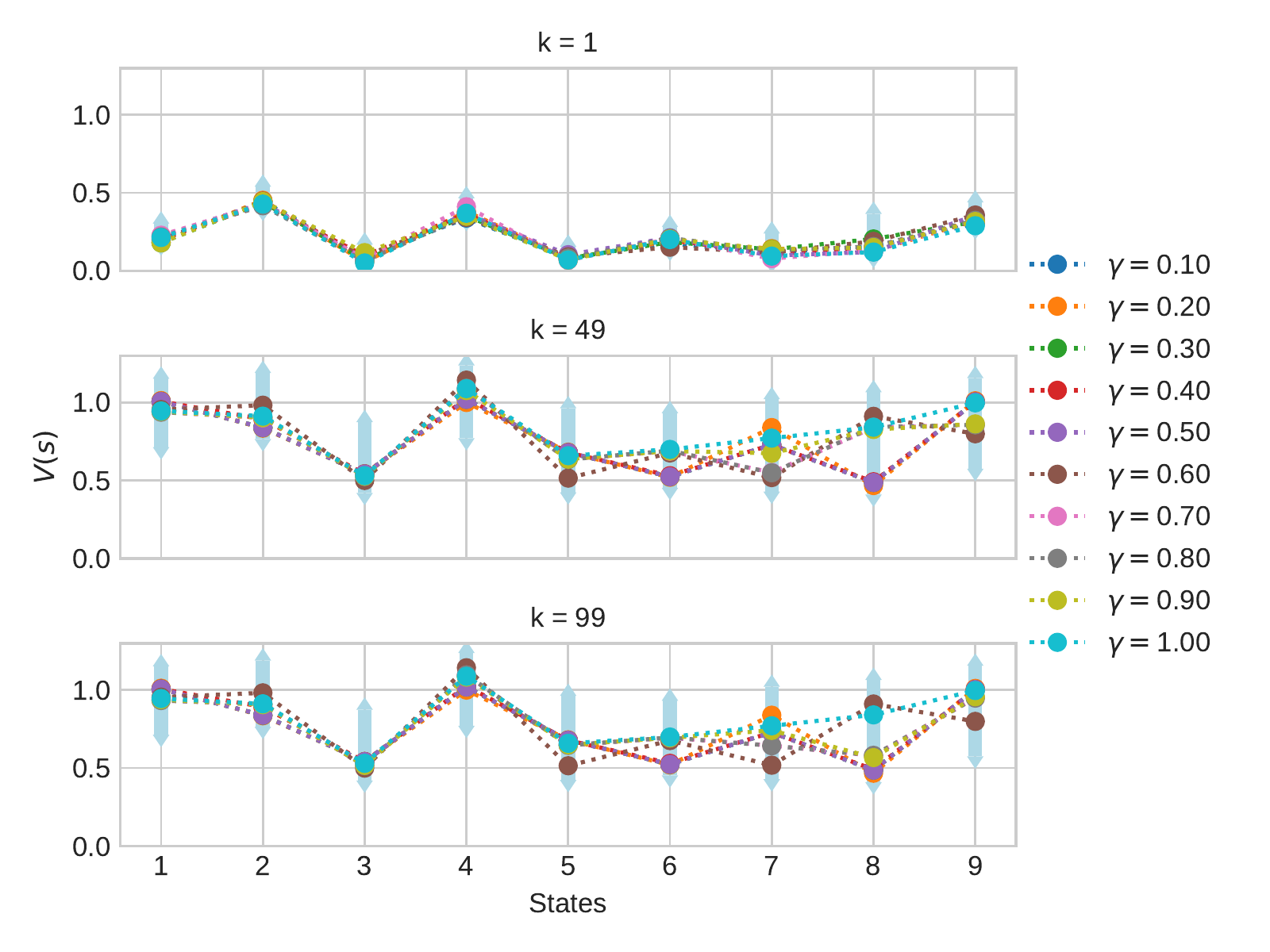}
    \caption{Player one's value function as a function of state at different iterations $k = \{1, 49, 99\}$. Range shown in blue is the bounded interval $\mc{V} = [\underline{V}^k, \overline{V}^k]$ at the corresponding iteration $k$. }
    \label{fig:discountFactor_VI}
\end{figure}

Suppose both players are updating their policies via value iteration~\eqref{eqn:optimalPol}. Player one performs value iteration with a discount factor of $\gamma_1 = 0.7$, while player two performs value iteration with an unknown discount factor $\gamma \in (0,1)$. Assuming both players' value functions are initialized to be $0$ in every state, we simulate player one's value function trajectories for different values of $\gamma$ in Figure~\ref{fig:discountFactor_VI}. 

Figure~\ref{fig:discountFactor_VI} shows that when player two utilizes different discount factors, player one experiences \emph{different} trajectories despite the fact that both players are utilizing value iteration to minimize their losses. However, the value function trajectory that player one follows is always bounded between the thresholds we derived from Proposition~\ref{prop:setConverge}. As Figure~\ref{fig:discountFactor_VI} shows, there does not seem to be any direct correlation between player two's discount factor and player one's value function. However, the interval bounds we derived do tightly approximate resulting value function trajectories.

\begin{figure}[ht]
    \centering
    \includegraphics[width=\columnwidth]{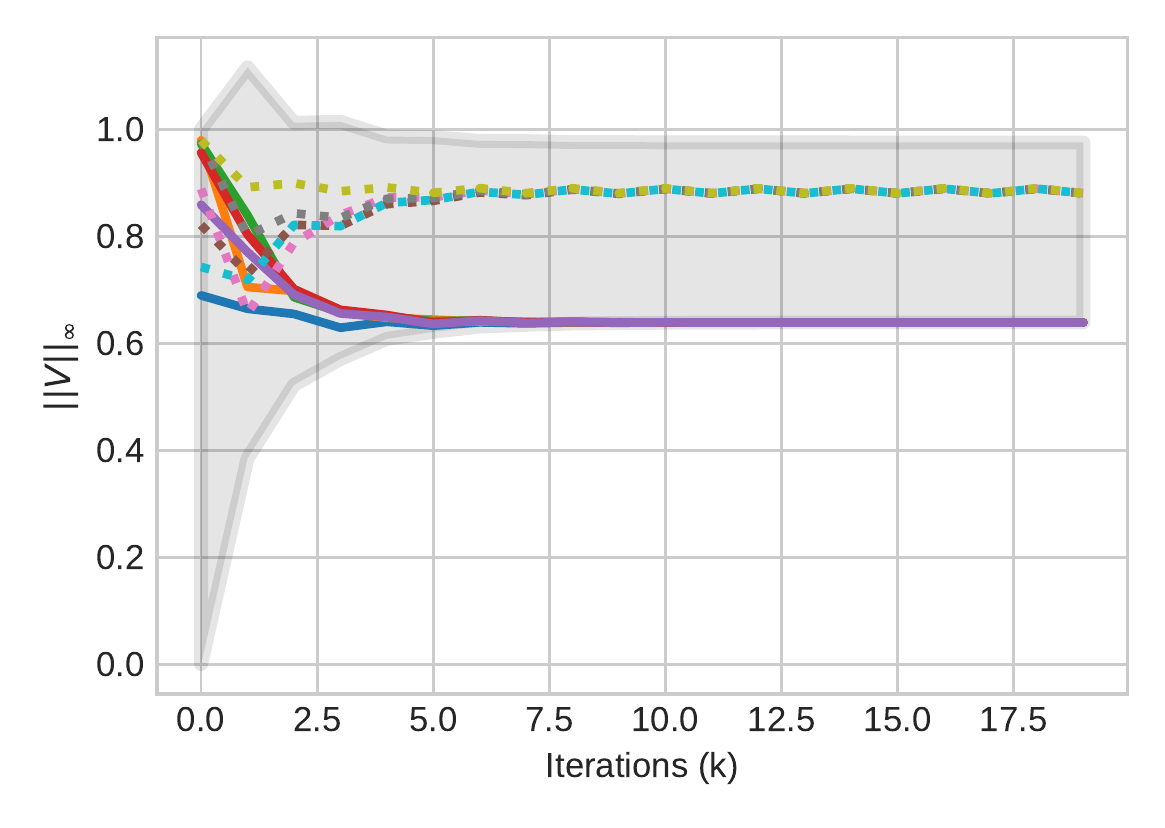}
    \caption{The infinity norm of player one's value functions as a function of iteration $k$.  }
    \label{fig:maxVsMin}
\end{figure}

Alternatively, suppose we know that player two has the same discount factor as player one, but we do not know player two's initial value function or if it is minimizing or maximizing its discounted objective. 
We analyze both scenarios: when player two is also minimizing its cost and when player two is maximizing its cost. In Figure~\ref{fig:maxVsMin}, the infinity norm of player one's value function at each iteration $k$ is shown with respect to these two scenarios. Both player one and player two's initial value function is randomly initialized as $V^0_s \in [0,1]$, $\forall \ s \in [9]$. Figure~\ref{fig:maxVsMin} plots player one's value function trajectory when player two utilizes value iteration towards different objectives: towards minimizing $C^2$ (player one's value functions shown in dotted lines) and towards maximizing $C^2$ (player one's value functions shown in solid lines). The grey region shows the predicted bounds as derived from Proposition~\ref{prop:setConverge}. 

As Figure~\ref{fig:maxVsMin} shows, player two's policy change causes a significant shift in player one's value function trajectory. When player two attempts to maximize its own cost parameter, player one's function achieves the absolute lower bound as predicted by Proposition~\ref{prop:intervalVIBounds}. This is due to the fact that at least four actions with different costs are available at each state. Since both players are only selecting from deterministic policies, they are bound to select different actions unless all actions have the exact same cost. On the other hand, if player two is minimizing its value function, then both players would precisely select the same state-actions at every time step. Then depending on the coupling matrix $A$, they may or may not choose a less costly action at the next step. This causes the limit cycle behaviour that the dotted trajectories exhibits. In terms of the tightness of the bounds we derived in Proposition~\ref{prop:setConverge}, we note that Figure~\ref{fig:maxVsMin} also shows the existence of trajectories which approach both the upper and lower bounds, therefore in practice the set-based bounds are shown to be tight. 


\section{Conclusion}
We have bounded the set of optimal value functions of the set-based Bellman operator associated with a discounted infinite horizon MDP. Our results are motivated by bounding optimal value functions of parameter uncertain MDPs and value functions trajectories of a player in a stochastic games. We demonstrate our example on a grid MDP and show that while player one's value function does not converge, the Hausdorff distance between the value function and the fixed point set of the set-based Bellman operator converges to zero. Future work includes extending the set-based analysis to consider uncertainty in the transition kernels to fully bound value function trajectories of learning algorithms in a general stochastic game. 



\bibliographystyle{plain}        
\bibliography{reference}           

\begin{thebibliography}{10}

\bibitem{abbad1992perturbation}
M.~{Abbad} and J.~A. {Filar}.
\newblock Perturbation and stability theory for markov control problems.
\newblock {\em IEEE Trans. Autom. Control}, 37(9):1415--1420, 1992.

\bibitem{accikmecse2015markov}
Beh{\c{c}}et A{\c{c}}{\i}kme{\c{s}}e and David~S Bayard.
\newblock Markov chain approach to probabilistic guidance for swarms of
  autonomous agents.
\newblock {\em Asian J. Control}, 17(4):1105--1124, 2015.

\bibitem{altman1994flow}
Eitan Altman.
\newblock Flow control using the theory of zero sum markov games.
\newblock {\em IEEE Trans. Autom. Control}, 39(4):814--818, 1994.

\bibitem{altman1993stability}
Eitan Altman and Vladimir~A Gaitsgory.
\newblock Stability and singular perturbations in constrained markov decision
  problems.
\newblock {\em IEEE Trans. Autom. Control}, 38(6):971--975, 1993.

\bibitem{ang2017game}
Samuel Ang, Hau Chan, Albert~Xin Jiang, and William Yeoh.
\newblock Game-theoretic goal recognition models with applications to security
  domains.
\newblock In {\em Int. Conf. Decision Game Theory Secur.}, pages 256--272.
  Springer, 2017.

\bibitem{bellemare2019geometric}
Marc Bellemare, Will Dabney, Robert Dadashi, Adrien~Ali Taiga, Pablo~Samuel
  Castro, Nicolas Le~Roux, Dale Schuurmans, Tor Lattimore, and Clare Lyle.
\newblock A geometric perspective on optimal representations for reinforcement
  learning.
\newblock In {\em Adv. Neural Inf. Process. Syst.}, pages 4358--4369, 2019.

\bibitem{bielecki1991singularly}
Tomasz~R Bielecki and Jerzy~A Filar.
\newblock Singularly perturbed markov control problem: Limiting average cost.
\newblock {\em Ann. Op. Res.}, 28(1):153--168, 1991.

\bibitem{bu2008comprehensive}
Lucian Bu, Robert Babu, Bart De~Schutter, et~al.
\newblock A comprehensive survey of multiagent reinforcement learning.
\newblock {\em IEEE Trans. Syst. Man Cybern. Part C (Appl. Rev.)},
  38(2):156--172, 2008.

\bibitem{chatterjee2004nash}
Krishnendu Chatterjee, Rupak Majumdar, and Marcin Jurdzi{\'n}ski.
\newblock On nash equilibria in stochastic games.
\newblock In {\em Int. Workshop Comput. Sci. Log.}, pages 26--40. Springer,
  2004.

\bibitem{chavent2004hausdorff}
Marie Chavent.
\newblock A hausdorff distance between hyper-rectangles for clustering interval
  data.
\newblock In {\em Classification, clustering, and data mining applications},
  pages 333--339. Springer, 2004.

\bibitem{DBLP:conf/icml/DadashiBTRS19}
Robert Dadashi, Marc~G. Bellemare, Adrien~Ali Taïga, Nicolas~Le Roux, and Dale
  Schuurmans.
\newblock The value function polytope in reinforcement learning.
\newblock In {\em Int. Conf. Machine Learning}, pages 1486--1495, 2019.

\bibitem{delage2010percentile}
Erick Delage and Shie Mannor.
\newblock Percentile optimization for markov decision processes with parameter
  uncertainty.
\newblock {\em Op. Res.}, 58(1):203--213, 2010.

\bibitem{demir2015decentralized}
Nazl{\i} Demir, Utku Eren, and Beh{\c{c}}et A{\c{c}}ikme{\c{s}}e.
\newblock Decentralized probabilistic density control of autonomous swarms with
  safety.
\newblock {\em Auton. Robots}, 39(4):537 --554, 2015.

\bibitem{dick2014online}
Travis Dick, Andras Gyorgy, and Csaba Szepesvari.
\newblock Online learning in markov decision processes with changing cost
  sequences.
\newblock In {\em Int. Conf. Machine Learning}, pages 512--520, 2014.

\bibitem{eisentraut2019stopping}
Julia Eisentraut, Jan K{\v{r}}et{\'\i}nsk{\`y}, and Alexej Rotar.
\newblock Stopping criteria for value and strategy iteration on concurrent
  stochastic reachability games.
\newblock {\em arXiv preprint arXiv:1909.08348}, 2019.

\bibitem{el2018controlled}
Mahmoud El~Chamie, Yue Yu, Beh{\c{c}}et A{\c{c}}{\i}kme{\c{s}}e, and Masahiro
  Ono.
\newblock Controlled markov processes with safety state constraints.
\newblock {\em IEEE Trans. Autom. Control}, 64(3):1003--1018, 2018.

\bibitem{eldosouky2016single}
Abdel~Rahman Eldosouky, Walid Saad, and Dusit Niyato.
\newblock Single controller stochastic games for optimized moving target
  defense.
\newblock In {\em 2016 IEEE Int. Conf Commun.}, pages 1--6. IEEE, 2016.

\bibitem{filar2012competitive}
Jerzy Filar and Koos Vrieze.
\newblock {\em Competitive Markov decision processes}.
\newblock Springer Science \& Business Media, 2012.

\bibitem{ganzfried2009computing}
Sam Ganzfried and Tuomas Sandholm.
\newblock Computing equilibria in multiplayer stochastic games of imperfect
  information.
\newblock In {\em 21st Int. Joint Conf. Artif. Intell.}, 2009.

\bibitem{givan2000bounded}
Robert Givan, Sonia Leach, and Thomas Dean.
\newblock Bounded-parameter markov decision processes.
\newblock {\em Artif. Intell.}, 122(1-2):71--109, 2000.

\bibitem{haddad2018interval}
Serge Haddad and Benjamin Monmege.
\newblock Interval iteration algorithm for mdps and imdps.
\newblock {\em Theor. Comput. Sci.}, 735:111--131, 2018.

\bibitem{henrikson1999completeness}
Jeff Henrikson.
\newblock Completeness and total boundedness of the hausdorff metric.
\newblock In {\em MIT Undergraduate J. Math.} Citeseer, 1999.

\bibitem{hu2003nash}
Junling Hu and Michael~P Wellman.
\newblock Nash q-learning for general-sum stochastic games.
\newblock {\em J. Mach. Learn. Res.}, 4(Nov):1039--1069, 2003.

\bibitem{iyengar2005robust}
Garud~N Iyengar.
\newblock Robust dynamic programming.
\newblock {\em Math. Op. Res.}, 30(2):257--280, 2005.

\bibitem{kearns2000fast}
Michael Kearns, Yishay Mansour, and Satinder Singh.
\newblock Fast planning in stochastic games.
\newblock In {\em Proc. 16th Conf. Uncertainty Artif. Intell.}, pages 309--316.
  Morgan Kaufmann Publishers Inc., 2000.

\bibitem{li2019tolling}
Sarah H.~Q. Li, Yue Yu, Daniel Calderone, Lillian Ratliff, and Behcet Acikmese.
\newblock Tolling for constraint satisfaction in markov decision process
  congestion games.
\newblock In {\em Amer. Control Conf.}, pages 1238--1243. IEEE, 2019.

\bibitem{li2020setBellman}
Sarah~H.Q. Li, Assal\'{e} Adj\'{e}, Pierre-Lo\"{i}c Garoche, and Beh{\c{c}}et
  A{\c{c}}{\i}kme{\c{s}}e.
\newblock Fixed points of the set-based bellman operator.
\newblock {\em arXiv preprint arXiv:2001.04535}, 2020.

\bibitem{littman1994markov}
Michael~L Littman.
\newblock Markov games as a framework for multi-agent reinforcement learning.
\newblock In {\em Mach. Learn. Proc. 1994}, pages 157--163. Elsevier, 1994.

\bibitem{littman2001value}
Michael~L Littman.
\newblock Value-function reinforcement learning in markov games.
\newblock {\em Cogn. Sys. Res.}, 2(1):55--66, 2001.

\bibitem{moore1966interval}
Ramon~E Moore.
\newblock {\em Interval analysis}, volume~4.
\newblock Prentice-Hall Englewood Cliffs, NJ, 1966.

\bibitem{prasad2015two}
HL~Prasad, Prashanth LA, and Shalabh Bhatnagar.
\newblock Two-timescale algorithms for learning nash equilibria in general-sum
  stochastic games.
\newblock In {\em Proc. 2015 Int. Conf. Auton. Agents Multiagent Sys.}, pages
  1371--1379. International Foundation for Autonomous Agents and Multiagent
  Systems, 2015.

\bibitem{puterman2014markov}
Martin~L Puterman.
\newblock {\em Markov Decision Processes.: Discrete Stochastic Dynamic
  Programming}.
\newblock John Wiley \& Sons, 2014.

\bibitem{rudin1964principles}
Walter Rudin et~al.
\newblock {\em Principles of mathematical analysis}, volume~3.
\newblock McGraw-hill New York, 1964.

\bibitem{shapley1953stochastic}
Lloyd~S Shapley.
\newblock Stochastic games.
\newblock {\em Proc. Nat. Acad. Sci.}, 39(10):1095--1100, 1953.

\bibitem{shiva2010game}
Sajjan Shiva, Sankardas Roy, and Dipankar Dasgupta.
\newblock Game theory for cyber security.
\newblock In {\em Proc. Sixth Annu. Workshop Cyber Secur. Inf. Intell. Res.},
  pages 1--4, 2010.

\bibitem{wei2017online}
Chen-Yu Wei, Yi-Te Hong, and Chi-Jen Lu.
\newblock Online reinforcement learning in stochastic games.
\newblock In {\em Adv. Neural Inf. Process. Sys.}, pages 4987--4997, 2017.

\bibitem{wiesemann2013robust}
Wolfram Wiesemann, Daniel Kuhn, and Ber{\c{c}} Rustem.
\newblock Robust markov decision processes.
\newblock {\em Math. Op. Res.}, 38(1):153--183, 2013.

\end{thebibliography}



\end{document}